\documentclass[preprintnumbers,amsmath,amssymb,floatfix,10pt,prd,twocolumn,
superscriptaddress,nofootinbib]{revtex4}
\usepackage{float}
\usepackage[parfill]{parskip}    
\usepackage{graphicx}
\usepackage{amssymb}
\usepackage{epstopdf}
\usepackage{color}
\usepackage{float}
\usepackage{amsmath}
\usepackage{orcidlink}
\usepackage{latexsym}
\usepackage{amsmath}
\usepackage{amssymb}
\usepackage{amsfonts}
\usepackage{blindtext}
\usepackage{subfigure}
\usepackage{xcolor}

\begin{document}

\title{{Kaniadakis Holographic Dark Energy: UV–IR Duality, Horizon Thermodynamics and Cosmological Constraints}
}

\author{Manuel Gonzalez-Espinoza \orcidlink{0000-0003-0961-8029}}
\email{manuel.gonzalez@upla.cl}
\affiliation{Laboratorio de investigaci\'on de C\'omputo de F\'isica, Facultad de Ciencias Naturales y Exactas, Universidad de Playa Ancha, Subida Leopoldo Carvallo 270, Valpara\'iso, Chile}

\author{Samuel Lepe}
\email{samuel.lepe@pucv.cl}

\affiliation{
Instituto de Física, Pontificia Universidad Católica de Valparaíso, Casilla 4950, Valparaíso, Chile.}

\author{Joel F. Saavedra \orcidlink{0000-0002-1430-3008}}
\email{joel.saavedra@pucv.cl}
\affiliation{
Instituto de Física, Pontificia Universidad Católica de Valparaíso, Casilla 4950, Valparaíso, Chile.}

\author{Francisco Tello-Ortiz \orcidlink{0000-0002-7104-5746}}
\email{francisco.tello@ufrontera.cl}
\affiliation{Departamento de Ciencias Físicas, Universidad de La Frontera, Casilla 54-D, 4811186 Temuco, Chile.}

\begin{abstract}
We investigate the cosmological and thermodynamic implications of holographic dark energy derived from the Kaniadakis deformation of the Bekenstein--Hawking entropy. In a spatially flat FLRW universe, the generalized entropy generates an effective dark-energy density with an infrared correction proportional to $H^{-2}$, naturally producing an ultraviolet--infrared structure in the cosmological dynamics. Using the Hayward--Kodama formulation of apparent-horizon thermodynamics, we derive a geometric equation of state and identify Van der Waals-type criticality, characterized by an inverted first-order phase transition and an inverted swallowtail Gibbs potential. We interpret this behavior as the thermodynamic signature of a geometric phase transition of the apparent horizon rather than standard fluid coexistence. We also consider an extended model with a $\dot H$ contribution inspired by the Granda--Oliveros cutoff and show that this unconventional critical behavior persists. At the perturbative level, the same ultraviolet--infrared competition controls the effective sound speed and dark-energy response through a dimensionless coupling $A(z)$, which acts as a geometric order parameter. In the minimal model, $A(z)$ may cross zero during matter domination, marking a transition between infrared- and ultraviolet-dominated regimes, whereas the extended scenario remains ultraviolet dominated. Finally, we confront both models with cosmic chronometers, PantheonPlus Type Ia supernovae, DESI DR2 baryon acoustic oscillations, and redshift-space distortion data. Bayesian comparison with $\Lambda$CDM, CPL, and standard holographic dark energy strongly disfavors the minimal model. The extended model yields the best maximum-likelihood fit among those considered, but remains disfavored relative to $\Lambda$CDM after accounting for its larger parameter volume.
\end{abstract}



\maketitle

\section{Introduction}

Understanding the physical origin of dark energy and the mechanism driving late-time cosmic acceleration remains one of the most profound challenges in contemporary cosmology. Although the $\Lambda$CDM model provides an excellent phenomenological description of observational data---including CMB anisotropies \cite{Planck2018}, baryon acoustic oscillations \cite{Alam2021}, and Type Ia supernovae \cite{Riess2022}---it leaves unresolved deep theoretical tensions. The cosmological constant problem highlights the enormous discrepancy between quantum vacuum energy estimates and the observed value \cite{Weinberg1989}, while the coincidence problem questions why dark energy and matter densities are of the same order precisely at the present epoch \cite{Zlatev1999}. These conceptual issues strongly suggest that cosmic acceleration may reflect new gravitational or thermodynamic physics rather than a finely tuned vacuum energy.

Holographic dark energy (HDE) offers a compelling framework rooted in quantum gravity principles. Originally proposed by Li \cite{Li2004} and based on the holographic principle \cite{Hooft1993,Susskind1995,Nojiri:2005pu,Nojiri:2017opc,Nojiri:2021iko,Nojiri:2022dkr,Nojiri:2023bom}, HDE relates the dark energy density to an infrared cutoff $L$ through $\rho_{\rm de} \propto L^{-2}$, using the Bekenstein--Hawking entropy relation $S=A/4$ \cite{Bekenstein1973,Hawking1975}. This construction naturally links ultraviolet and infrared scales and yields an effective equation of state parameter close to $w\simeq -1$, in agreement with observations.

However, the classical area law is expected to receive corrections from quantum, statistical, or gravitational effects near Planck scales. This expectation has motivated the exploration of generalized entropy formalisms, including Tsallis \cite{Tsallis1988,Nojiri:2022aof}, Rényi \cite{Renyi1961}, Barrow \cite{Barrow2020,Nojiri:2021jxf}, and Kaniadakis entropies \cite{Kaniadakis2001,Kaniadakis2002}. In \cite{Hernandez-Almada:2021aiw} they performed an important observational and dynamical-systems study of Kaniadakis holographic dark energy. Using cosmic chronometers, Type Ia supernovae and BAO data, a Markov Chain Monte Carlo analysis showed that the Kaniadakis deformation parameter $K$ is constrained to values near zero, the limit in which it $S_K$ recovers the Bekenstein-Hawking entropy, and that the Universe exhibits a matter dominated past attractor and a dark energy dominated future attractor. In the present manuscript, we use the latest DESI DR2 BAO dataset and the PantheonPlus SNe~Ia compilation, and   we include RSD growth-rate measurements and also, we provide a complete thermodynamic and perturbative analysis of the model, which was not considered in \cite{Hernandez-Almada:2021aiw}.
Among these, the Kaniadakis entropy is particularly well suited for relativistic systems, as it emerges from $\kappa$-deformed statistics consistent with Lorentz symmetry and introduces a single deformation parameter $K$ while preserving a consistent thermodynamic structure.
When implemented holographically in a cosmological setting\footnote{See \cite{Drepanou2022,Moradpour2020} for related cosmological analyses in the Kaniadakis setting.}, the Kaniadakis entropy modifies the standard HDE density by introducing an additional contribution proportional to $K^2 H^{-2}$. This correction generates enhanced infrared sensitivity, significantly affecting both the background expansion and the effective dark energy equation of state $w_{\rm de}(z)$. More importantly, it modifies the thermodynamic properties of the apparent horizon—the natural causal boundary in FLRW cosmology—through its influence on the Hayward--Kodama surface gravity \cite{Kodama1980,Hayward1998} and the Misner--Sharp quasi-local energy \cite{MisnerSharp1964}.

The thermodynamic interpretation of gravity provides the natural theoretical framework to investigate these modifications. Following the foundational works of Bekenstein and Hawking \cite{Bekenstein1973,Hawking1975}, Jacobson demonstrated that Einstein’s equations can be derived from the Clausius relation $\delta Q = T\delta S$ applied to local Rindler horizons \cite{Jacobson:1995ab}. Complementary formulations based on the unified first law and apparent horizon thermodynamics were developed by Hayward and others \cite{Hayward1998}. These results establish that the interplay between spacetime geometry and thermodynamics is not accidental but reflects a fundamental property of gravitational dynamics.

From an observational standpoint, any viable modification of holographic dark energy must confront high-precision late-time data. Cosmic chronometers provide direct reconstructions of $H(z)$ from differential galaxy ages \cite{Moresco2016,Ratsimbazafy2017}, the PantheonPlus sample delivers improved Type Ia supernova constraints over a broad redshift range \cite{Brout2022}, and the latest BAO measurements from DESI DR2 further tighten constraints on the expansion history \cite{DESI2024,DESI2024BAO}. {Recent DESI DR2 measurements have also been employed to test a broad
class of entropy-based cosmological scenarios, including Kaniadakis
holographic dark energy as well as Barrow-, Tsallis-, and other
generalized entropic models, providing increasingly stringent tests
of departures from the standard cosmological framework
\cite{Leizerovich:2026pfy,Luciano:2025ykr,Luciano:2025hjn}.}

Since entropic deformations typically manifest as infrared corrections, they become dynamically relevant at low redshift, although degeneracies with $\Lambda$CDM may persist at the homogeneous level \cite{Linder2005,Wang2008}. A consistent assessment therefore requires both thermodynamic consistency and systematic observational confrontation.

Motivated by these considerations, in this work we present a comprehensive analysis of Kaniadakis holographic cosmology. First, we derive the effective equation of state of the apparent horizon, $P=P(v,T)$, directly from the unified first law without invoking the quasi-static approximation, uncovering a nontrivial critical structure. Second, we contrast this framework with alternative infrared prescriptions, particularly the Granda-Oliveros cutoff \cite{GrandaOliveros2008}, which provides a local reformulation of HDE. Remarkably, {we find that the Kaniadakis deformation induces an inverted first-order phase transition whose Gibbs free energy develops an inverted swallowtail structure.} {A particularly novel feature of the present model is that this swallowtail does not describe a conventional fluid coexistence region, but rather the thermodynamic imprint of a geometric phase transition of the apparent horizon.}


{From this perspective, the model provides a unified geometric interpretation in which the entropic deformation is reflected consistently across background dynamics, horizon thermodynamics, and structure growth.}

The manuscript is organized as follows. In Sect.~\ref{sec2} the holographic dark energy model is presented. The Sect.~ \ref{sec3} introduces the apparent horizon dynamics and in Sect.~\ref{sec4} presents the equation of state. Sect.~ \ref{sec5} is devoted to a detailed thermodynamic analysis of the system, by imposing the standard criticality conditions. In Sect.~\ref{sec6} we investigate the classical stability of the model through the effective sound speed. Sect.~\ref{sec7} derives the modified background dynamics and obtains explicit expressions for the Hubble expansion rate. In Sect.~ \ref{sec8} we study cosmological perturbations in the Newtonian gauge. These results provide the basis for the observational analysis presented in Sect.~ \ref{sec9}, where the model is confronted with late-time cosmological probes through a joint statistical analysis employing cosmic chronometers, the PantheonPlus Type Ia supernova sample, and baryon acoustic oscillation measurements from DESI. Finally, Sect.~ \ref{sec10} summarizes the main results and discusses their implications for holographic cosmology. Technical derivations and complementary analyzes are presented in Appendices \ref{app:A}, \ref{app:B} and \ref{app:C}.

Throughout the manuscript the signature is $\{-,+,+,+\}$, and units where $8\pi G=c=1$.

\section{Kaniadakis entropy and its holographic formulation}\label{sec2}

The statistical foundation of the Kaniadakis entropy stems from a relativistic generalization of Boltzmann Gibbs theory \cite{Kaniadakis2001,Kaniadakis2002}. Unlike other non-additive entropies such as Tsallis or Rényi, the Kaniadakis deformation preserves fundamental structures of special relativity by introducing a $\kappa$-deformed exponential and logarithm constructed to remain invariant under Lorentz transformations. The entropy functional
\begin{equation}
    S_{K} = -\int d\Gamma\, f \ln_{\kappa} f ,
\end{equation}
where $\ln_{\kappa}$ denotes the Kaniadakis logarithm, produces a well-defined thermodynamic framework with a single deformation parameter $K$ and smoothly reduces to the Boltzmann--Gibbs entropy in the limit $K\to 0$. This property makes the Kaniadakis formalism particularly attractive for gravitational
systems, where high-energy corrections and non-Gaussian statistics naturally arise.
As shown in \cite{Moradpour2020}, the Kaniadakis entropy may be applied to black holes and cosmological horizons by assuming that the horizon microstates follow a $\kappa$-deformed statistical distribution. In this case, the entropy becomes a function of the Bekenstein-Hawking entropy given by $S_{h}=A/4$, where $A$ is the apparent horizon area. Then, 
\begin{equation}
S_{K}(A)= \frac{1}{K}\,\sinh\!\left(K\,S_{h}\right),
\label{SK-A}
\end{equation}
For small deformations ($K S_{h}\ll 1$), Eq.~\eqref{SK-A} expands to
\begin{equation}\label{expansion1}
S_{K}(S_{h})
= S_{h}+\frac{1}{6}K^2 S^{3}_{h}
+ \mathcal{O}(K^{4}),
\end{equation}

which clearly shows that the leading term reproduces the Bekenstein-Hawking entropy \cite{Hawking1975,Bekenstein1973}. This unveils that the Kaniadakis modification contributes higher-order geometric corrections to the standard area law.
In holographic dark energy models, the energy density is directly linked to the entropy associated with the infrared cutoff length $L$ \cite{Li:2004rb,Wang:2016och}.
In the context of horizon thermodynamics, a modified entropy immediately implies a modified energy associated with the apparent horizon.

Using the holographic dark energy argument \cite{Li:2004rb,Wang:2016och,Moradpour:2020dfm}, that is, 
\begin{equation}
    \rho_{de}L^{4} \leq S_{h},
\end{equation}

and using the expansion \eqref{expansion1} evaluated at the apparent horizon area $A$, one arrives at
\begin{equation}
\rho_{\text{de}}L^{4}\leq \frac{A}{4}+\frac{1}{6}K^{2}\left( \frac{A}{4}\right) ^{3}, 
\quad \mbox{with}\quad \frac{A}{4}=\pi L^{2}, 
\label{eq:2.4}
\end{equation}
then
\begin{equation}
\rho_{\text{de}}L^{4} \leq \pi L^{2}+\frac{\pi ^{3}}{6}K^{2}L^{6}
\Longrightarrow 
\rho_{\text{de}}\leq \pi L^{-2}+\frac{\pi ^{3}}{6}K^{2}L^{2}.\label{eq6}
\end{equation}


Now, considering $L=H^{-1}$, one obtains the HDE model
\begin{equation}\label{eq7}
    \rho
_{\text{de}}=3c^{2} H^{2}+\frac{K^{2}}{ H^{2}}.
\end{equation}

Some clarification of terms is in order. The UV--IR structure found here, namely the presence of the term $3c^2H^2$ accounting for the ultraviolet Bekenstein-Hawking part and the term $K^2/H^2$ describing the infrared correction arising from the Kaniadakis deformation, should not be mistaken with the UV--IR duality in the strict field theoretical sense of Maldacena-Susskind-Witten \cite{Cohen:1998zx}. In the original holographic dark energy proposal of Li~\cite{Li:2004rb} the infrared cutoff $L$ and the ultraviolet scale are connected through the entropy bound, thus establishing a UV-IR connection at the level of the energy density. The Kaniadakis deformation converts this single-scale structure into a two-scale setting: the leading term $3c^2H^2\propto S_{\rm BH}$ is determined by the Hubble rate as a UV quantity while the subleading correction $K^2/H^2\propto S_{\rm BH}^3$ (arising from the third-order term in expansion (3)) introduces a different infrared sensitivity. The resulting functional form $\rho_{\rm de}=\alpha H^2+\beta H^{-2}$ has been recently obtained also from a distinct two-parameter entropic functional~\cite{Luciano:2026eiy}, which suggests that this two-scale structure is a generic feature of holographic scenarios built on generalized entropies, rather than an idiosyncrasy of the Kaniadakis $\kappa$-statistics. In this manuscript the expression ``UV-IR duality'' will thus refer specifically to this competition between the two scales encoded in $\rho_{\rm de}$. It should not be read as a claim of gauge/gravity duality in the sense of AdS/CFT.
The expression in Eq.~(\ref{eq7}) reveals in a transparent way how the
Kaniadakis deformation parameter $K$ modifies the standard holographic
dark energy prescription. In particular, while the usual term
$\rho_{\text{de}}\sim H^{2}$ arises from the leading Bekenstein-Hawking
entropy contribution, the additional term proportional to $K^{2}H^{-2}$
originates from the higher-order correction in the entropy expansion.
This correction effectively introduces an infrared contribution to the
energy density, which becomes dominant in the late-time regime where
$H\to 0$. As a consequence, the Kaniadakis parameter induces a natural
self-accelerating behavior without the need of an explicit cosmological
constant. From a physical perspective, this shows that the deformation
of the entropy-area law propagates non-trivially into the cosmological
dynamics, generating a two-scale structure in $\rho_{\text{de}}$:
a UV-like term controlled by $H^{2}$ and an IR-like term controlled by
$H^{-2}$. 

 \section{The apparent horizon dynamics}\label{sec3}

Let's start by considering a spatially flat ($k=0$) FLRW Universe,
\begin{equation}
ds^2 = -dt^2 + a^2(t)\left[dr^2 + r^2 d\Omega^2\right],
\end{equation}
where $a(t)$ is the scale factor and $d\Omega^{2}=d\theta^{2}+\sin^{2}\theta d\varphi^{2}$.
For the above line element, the Friedmann field equations read 
\begin{eqnarray}\label{eq9}
3H^{2} &=&\rho +\rho _{\text{de}},  \\ \label{eq10}
\dot{H} &=&-\frac{1}{2}\left(\rho +p+\rho _{\text{de}}+p_{\text{de}}\right),
\end{eqnarray}
along with the conservation equations
\begin{equation}\label{conservation}
    \dot{\rho}+3H\left( \rho +p\right) =0,\quad\dot{\rho}%
_{\text{de}}+3H\left( \rho _{\text{de}}+p_{\text{de}}\right) =0,
\end{equation}
with Hubble given as usual by $H=\dot{a}/a$.
It is worth mentioning that, the components of the Universe for the present case are a perfect fluid matter distribution characterized by $\{\rho,p\}$ and the dark sector described by $\{\rho_{\text{de}},p_{\text{de}}\}$. Moreover, each sector is separately conserved, meaning that there is not energy exchange between them.

A spatially flat, $k=0$, FLRW Universe has as causal boundary an apparent horizon $R_{A}$ coinciding with the Hubble's horizon \cite{Faraoni:2011hf}, that is,
\begin{equation}\label{eq12}
    R_{A}=\frac{1}{H},
\end{equation}
The apparent horizon $R_{A}$ is the surface satisfying the following equation
\begin{equation}\label{EqAH}
    h^{ij}\nabla_{i}R\nabla_{j}R\bigg|_{R=R_{A}}=0, \quad i,j=t,r,
\end{equation}
where $R$ is the areal radius defined as $R(t,r)=a(t)r$, being  $r$ the co-moving radial distance and $h^{ij}$ the inverse metric of the two--dimensional metric $h_{ij}$ on the $t$-$r$ plane. These definitions come from the FLRW line element 
re-expressed using the warped product as \cite{Faraoni:2011hf}
\begin{equation}\label{warped}
    ds^{2}=h_{ij}dx^{i} dx^{j}+R^{2}d \Omega^2.
\end{equation}
As we are going to explore the thermodynamics behavior of this corrected holographic dark energy model, to do so one needs to ensure that all geometric properties can be properly interpreted as thermodynamic variable states. In this regards, to perform a thermodynamics study on the apparent horizon it is necessary to introduce a suitable definition for the surface gravity. Since the apparent horizon is a dynamical object, the surface gravity leading to a temperature definition in this case is the well-known Hayward-Kodama surface gravity given by \cite{Kodama:1979vn,Hayward:1997jp}
\begin{equation}
    \kappa = \frac{1}{2\sqrt{-h}}\partial_{i}(\sqrt{-h}h^{ij}\partial_{j}R),
\end{equation}
yielding 
\begin{equation}
    \kappa = -\frac{R}{2}\left(\dot{H}+2H^{2} \right).
\end{equation}
With this expression, one can
obtain the temperature at the apparent horizon $R=R_{A}$ as follows
\begin{equation}
    T_{A} = -\frac{\kappa}{2\pi} = \frac{H}{2\pi}\left(1-\frac{\dot{R}_{A}}{2HR_{A}}\right). \label{eq:temp}
\end{equation}
The minus sign in front the surface gravity, ensure a positive temperature. This is consistent with the fact that the apparent horizon corresponds to an inner-past horizon, since $\kappa<0$ \cite{Cai:2006rs}.

 \section{Modified holographic dark energy model and the equation of state}\label{sec4}

 Following the approach given in \cite{Cruz:2023xjp}, the Friedmann Eq. (\ref{eq9}) adopts the following form 
 \begin{equation}\label{eq119}
{3H^{2}}=\left( 1+r\right) \left( 3c^{2}H^{2}+\frac{K^{2}}{H^{2}}\right).  
\end{equation}
Here, $r$ is the so-called coincident parameter defined by $r\equiv \rho/\rho_{\text{de}}$. The introduction of this parameter is helpful because through it one can check the impact introduced by the dark sector into the thermodynamic description. Now using Eq. (\ref{eq12}), the expression (\ref{eq9}) can be recast as 
\begin{equation}\label{eq20}
1+r=\frac{1}{c^{2}+\frac{K^{2}}{3} R_{A}^{4}}.
\end{equation}

The dynamical nature of the $R_{A}$ suggests that the thermodynamic first law should be extended as proposed in \cite{Hayward:1997jp}.
The main ingredients of this formulation are: i) the Misner--Sharp energy, ii) the  Hayward-Kodama surface gravity and iii) the areal volume $V_{A}=\frac{4\pi}{3}R^{3}_{A}$. Additionally, it is necessary to introduce new quantities to reach the desired thermodynamics description. These quantities are: iv) the  work density $W$ and the v) energy--flux $\psi_{i}$ across $R_{A}$. The definition of these new pieces is given by
\begin{equation}\label{workdensity}
    W\equiv-\frac{1}{2}h_{ij}T^{ij},
\end{equation}
and
\begin{equation}
    \psi_{i}\equiv T^{j}_{i}\nabla_{i}R+W\nabla_{i}R,
\end{equation}
respectively. Furthermore, $T^{ij}$ represents the projected energy-momentum tensor components and $A\psi_{i}$ is so-called the energy-supply vector. 

Importantly, the work density is related with the pressure of the system. Concretely,
\begin{equation}\label{eq34}
P(R_{A},T)\equiv W=\frac{1}{2}\left(\rho-p\right),   \end{equation}
where the radius of the $R_{A}$ is connected with the 
 the specific volume as follows $v=2R_{A}$ \cite{Kubiznak:2012wp}. Providing, in this way a relation of the type
\begin{equation}\label{eq41}
    P=P(v,T),
\end{equation}
as usual. Putting together Eqs. (\ref{eq10}), (\ref{eq12}), (\ref{eq:temp}),  (\ref{eq20}) and (\ref{eq34}) one gets
\begin{equation}\label{eos1}
P=\frac{T_{A}}{v}+\frac{1}{2}\left[ \frac{K^{2}}{4}v^{2}-\frac{12\left(
1/3-c^{2}\right) }{v^{2}}\right]. 
\end{equation}
It is clear that from (\ref{eos1}) that $\{K,c\}=\{0,0\}$ leads to well-known result for pure Einstein gravity.

Another interesting proposal within the framework of holographic dark energy models, is the Granda-Oliveros (GO) proposal \cite{Granda:2008dk}. This provides a causal and fully local reformulation of holographic dark energy by replacing the traditional infrared cutoff-typically associated with the future event horizon with a scale constructed solely from the Hubble parameter and its time derivative. 

With this motivation at hand, we propose here the following holographic dark energy model, including both Kaniadakis and GO like contributions
\begin{equation}\label{eq45}
     \rho
_{\text{de}}=3c^{2}H^{2}+\frac{K^{2}}{H^{2}}+3\beta \dot{H}.
 \end{equation} Following a similar procedure, the equation of state (\ref{eq41}), in this case, leads to  
\begin{equation}\label{eos2}
P(v,T)=\left[ 1+\frac{3}{2}\beta \right] \frac{T_{A}}{v}+\frac{1}{2}\left[ \frac{
K^{2}}{4}v^{2}-\frac{12\left( 1/3-c^{2}+2\beta \right) }{v^{2}}\right].
\end{equation}

It is worth mentioning that, to obtain the above expression, the Friedmann equations \eqref{eq9} and \eqref{eq10} took into account the dynamical contribution given by GO like term. 

Now, we are in position to examine the thermodynamics of both models, \eqref{eos1} and \eqref{eos2}. This will be matter of the next section.

\section{Phase transitions and criticality}\label{sec5}

To study phase transitions, it is important to determine the possible existence of critical points. To do this, one appeals to the critical conditions given by 
\begin{equation}
    \left(\frac{\partial P}{\partial v}\right)\bigg{|}_{T_{A}}=\left(\frac{\partial^{2} P}{\partial v^{2}}\right)\bigg{|}_{T_{A}}=0.
\end{equation}
The solution of the above system leads to the critical volume and critical temperature. For the equation of state (\ref{eos1}) one obtains   
\begin{equation}
v_{c}=2\left( \frac{1/3-c^{2}}{K^{2}}\right) ^{1/4},
\end{equation}%
and 

\begin{equation}
T_{c}=8\left( 1/3-c^{2}\right) \left( \frac{K^{2}}{1/3-c^{2}}\right) ^{1/4}.
\end{equation}
A positive defined critical volume $v_{c}$ and critical temperature $T_{c}$ is possible if and only if the holographic parameter $c$ satisfies $c^{2}<\frac{1}{{3}}$.
Besides, the critical pressure is given by 
\begin{equation}
P_{c}=3\sqrt{1/3-c^{2}}K,  
\end{equation}
where the condition imposed on $c$ yields to a real critical pressure. For this system the compressibility factor $Z_{c}$ is
\begin{equation}
Z_{c}\equiv\frac{P_{c}v_{c}}{T_{c}}=2\frac{3}{8},
\end{equation}
that is, twice of the given for a genuine Van der Waals liquid-gas phase transition. On the other hand, when $\beta$ terms are present, from the equation of state (\ref{eos2}) the criticality conditions provide the following critical values,
\begin{equation}\label{eq36}
v_{c}=2\left( \frac{1/3-c^{2}+2\beta }{K^{2}}\right) ^{1/4}, 
\end{equation}
and 
\begin{equation}\label{eq37}
    T_{c} =8\sqrt{K}\left( 1/3-c^{2}+2\beta \right) ^{3/4}\left( 1+\frac{3}{2}
\beta \right) ^{-1}, 
\end{equation}
and
\begin{eqnarray}
P_{c} =3\sqrt{1/3-c^{2}+2\beta }K.
\end{eqnarray}
Here, $Z_{c}$ is more involved than in the previous case depending, on the $\beta$ parameter, that is,
\begin{equation}
    \frac{P_{c}v_{c}}{T_{c}} =2\left( 1+\frac{3}{2}\beta \right)\frac{3}{8}.
\end{equation}
 
\begin{figure}[H]
    \centering
\includegraphics[width=0.45\textwidth]{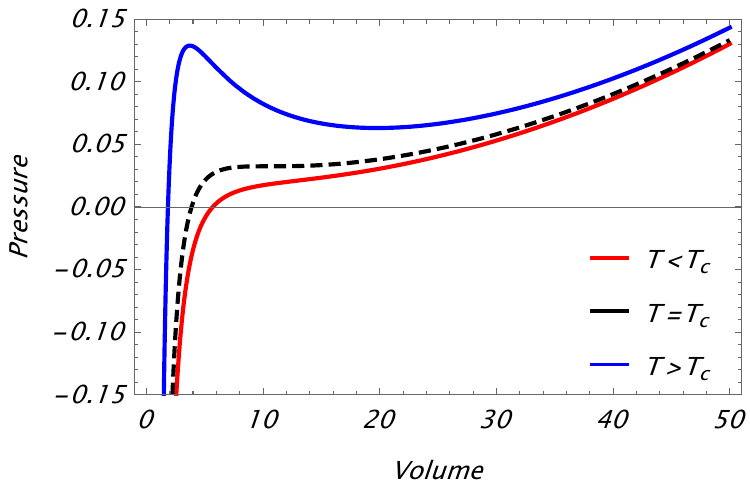}\ 
    \caption{The trend of the pressure (\ref{eos1}) versus the volume for the model \eqref{eq7}. The system presents an inverted first order phase transition above the critical temperature. This plot considers $\{c,K\}=\{0.2;0.02\}$. }
    \label{fig1}
\end{figure}

Now we are in position to explore the behavior of both, (\ref{eos1}) and (\ref{eos2}). For the former, in Fig. \ref{fig1} it is displayed the trend of the pressure against the volume. As can be observed, the model presents an unusual behavior, that is, the phase transition takes place for values of $T$ above the critical one (see blue line). Furthermore, the system stars taking small negative values for the pressure with small volumes and then grows with increasing volume up to a local maximum. This behavior is associated with unstable thermodynamic states, as the isothermal compressibility becomes negative. After reaching the local maximum, the pressure decreases with increasing volume. Here, the system presents an allowed stable thermodynamic state. Nevertheless, at some point the pressure again starts to increase in magnitude with increasing volume. This fact reveals that the system becomes unstable for volumes that are large enough. 

\begin{figure*}
    \centering
\includegraphics[width=0.43\textwidth]{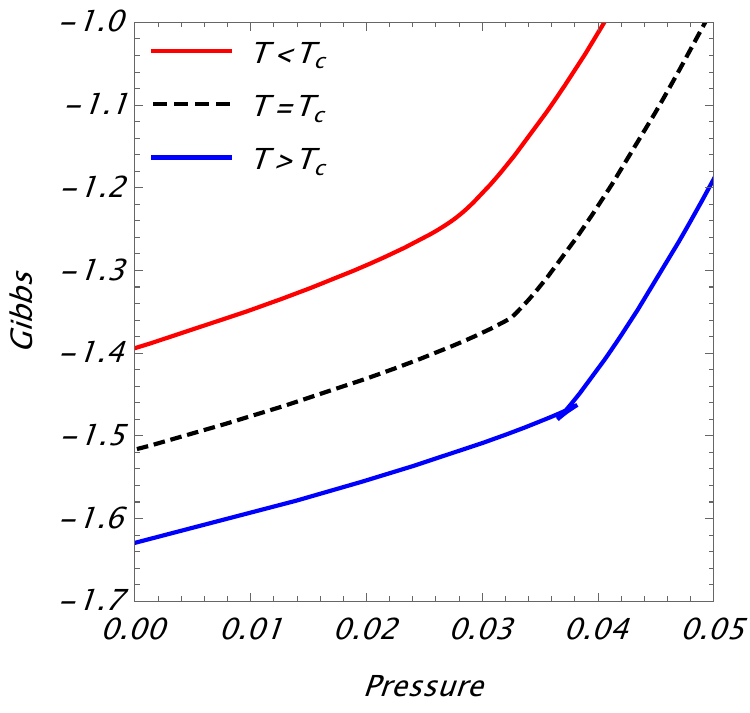}\, 
\includegraphics[width=0.44\textwidth]{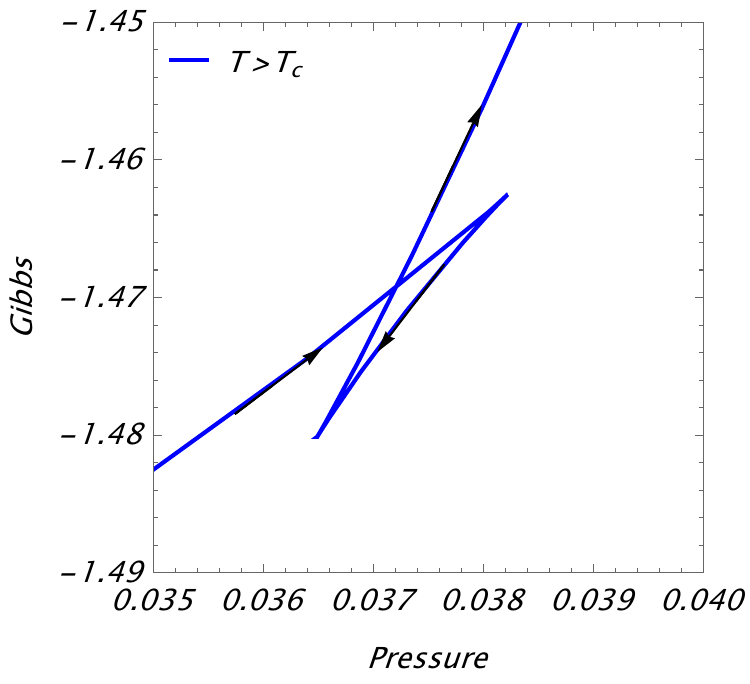}
    \caption{\textbf{Left panel}: The Gibbs free energy against the pressure for the model \eqref{eq45}. For those values below the critical temperature (red line), the Gibbs function is smooth, while increasing the temperature and crossing the critical one (blue line), develop an inverted swallowtail structure associated with a geometric phase transition of the apparent horizon shape accounting for a non-conventional first order phase transition. \textbf{Right panel}: A zoom of the swallowtail. It should be taking into account that the curve traversed from small to huge volumes. As the pressure increases with volume, the Gibbs free energy never reaches a minimum value as desired for thermodynamic stability.    }
    \label{fig2}
\end{figure*}

To further corroborate the above statements, in Fig. \ref{fig2} the trend of Gibbs free energy 
\begin{equation}\label{G1}    G(v,T)=\frac{4\left(3c^{2}-1\right)}{v}+\frac{v^{3}}{12}K^{2}-T\ln v,
\end{equation}
is depicted. The right panel of Fig. \ref{fig2} shows a zoom of the Gibbs free energy for $T>T_{c}$, where the system shows the shape of the inverted swallowtail.  This is so because the Gibbs free energy is not reaching its minimum value. In comparison with the usual first order phase transition presented by a real gas, here the usual unstable states are not forbidden for the system. 

{The inverted swallowtail structure can be interpreted as the thermodynamic signature of a geometric phase transition of the apparent horizon. Since all thermodynamic variables are constructed from the horizon radius and the associated Hayward-Kodama temperature, the critical behavior reflects a transition between two geometric regimes: a UV-dominated phase governed by the standard holographic term $3c^2H^2$ and an IR-dominated phase controlled by the Kaniadakis correction $K^2/H^2$. The perturbative coupling $A(z)$ (see below Sect.~\ref{sec8}) plays the role of an effective geometric order parameter, whose sign reversal identifies the transition between both regimes.  From this perspective, the swallowtail does not represent a conventional liquid-gas coexistence, but rather the thermodynamic imprint of a geometric reorganization of the apparent horizon and the onset of a non-equilibrium regime in its effective thermodynamic description.}

\begin{figure}[H]
    \centering
\includegraphics[width=0.45\textwidth]{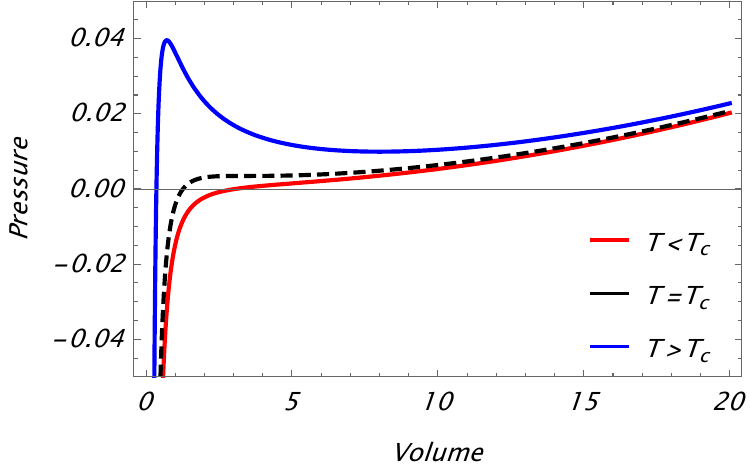}\ 
    \caption{The trend of the pressure (\ref{eos2}) versus the volume for the model \eqref{eq45}. The system presents an inverted first order phase transition above the critical temperature. This plot considers $\{c;K;\beta\}=\{0.6;0.02;0.015\}$. }
    \label{fig4}
\end{figure}

Next, we analyze the case for the equation of state (\ref{eos2}). In this case, one can obtain a relation between $c$ and $\beta$. Of course, for $\beta=0$ one recovers the previous model \eqref{eos1}. Besides, as in this model $c^{2}$ should be lees than unity, we can restrict $\beta$ considering this fact to have a positive critical volume and critical temperature. Therefore, one has
$\beta>(3c^{2}-1)/6$.
On the other hand, to further ensure $T_{c}>0$ it is necessary to enforce
$\beta>-2/3$.
If we restrict ourselves to the case $1/3<c^{2}<1$, then from $\beta$'s constraints, the second one can be discarded. The Fig. \ref{fig4} displays the pressure including the $\beta$ terms. It can be appreciated that there is an \emph{inverted} first order phase transition.  

\begin{figure}[H]
    \centering
\includegraphics[width=0.45\textwidth]{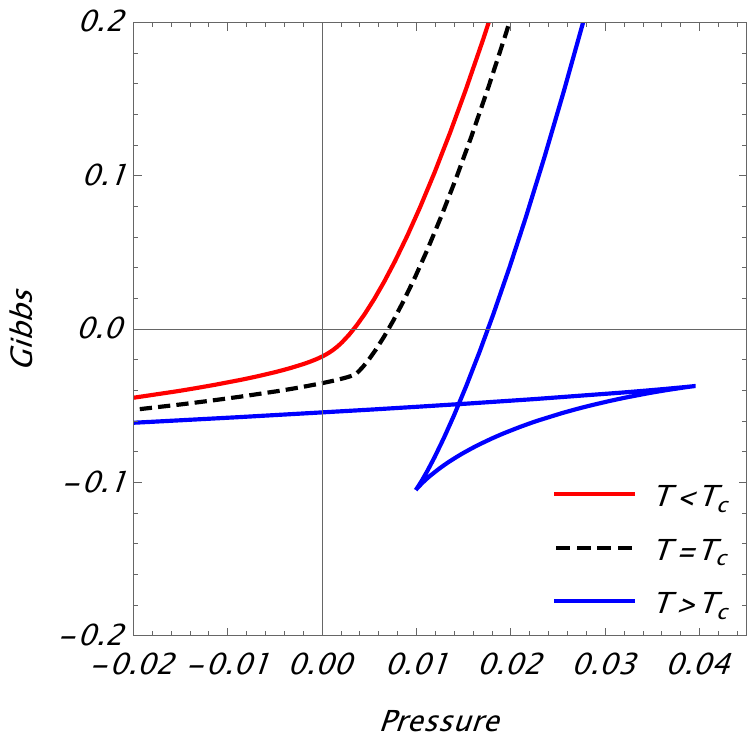}\ 
    \caption{The Gibbs free energy against the pressure for the model \eqref{eq7}. For those values below the critical temperature (red line), the Gibbs function is smooth, while increasing the temperature and crossing the critical one (blue line), develop an inverted swallowtail structure associated with a geometric phase transition of the apparent horizon shape accounting for a non-conventional first order phase transition.   }
    \label{fig5}
\end{figure}

For this case, Fig. \ref{fig5} shows the Gibbs free energy as a function of pressure.

\begin{equation}
\begin{split}
G(v,T)=\frac{1}{v}\bigg[-4+12 c^2+k^2 v^4/12-24 \beta &\\- (T v/2)(2+3 \beta) \ln v\bigg].
\end{split}
\end{equation}
As before, above the critical temperature the system presents a swallowtail shape which is non-physical.  

One basic question is whether the inverted swallowtail structure corresponds to a real geometric phase transition of the apparent horizon or simply to the particular functional form of the effective equation of state~(24). We favour the latter interpretation for the following reasons:
The thermodynamic variables that enter the equation of state, the specific volume $v=2R_A$, the Hayward-Kodama temperature $T_A$, and the work density $P$, are constructed entirely from the intrinsic geometry of the FLRW spacetime via the unified first law~\cite{Hayward:1997jp,Cai:2006rs}. They do not describe any external fluid or phenomenological matter model. Rather, they encode the response of the causal structure of the universe to changes in the horizon geometry. This is exactly the context where $P$--$V$ criticality has been shown to be physically meaningful for black holes in extended phase space~\cite{Kubiznak:2012wp}, where the Van der Waals analogy is recognized at the level of the geometry rather than a specific matter content.
In the present model, the inversion of the phase transition, its occurrence above $T_c$ rather than below, follows necessarily from the sign structure of the Kaniadakis correction. The term $K^2v^2/4$ in Eq.~(24) has the opposite sign to a standard Van der Waals equation of state, because the infrared contribution $K^2/H^2$ \emph{increases} with the expansion of the horizon (i.e. with $v$), rather than decreasing. This change of sign is not an arbitrary parameter of the model, but it is determined by the positive definite correction $K^2S_h^3/6$ of expansion~(3), which guarantees that the Kaniadakis entropy is always larger than the Bekenstein-Hawking one. The inverted swallowtail in the Gibbs free energy is hence a thermodynamic imprint of a geometric reorganization of the apparent horizon between a UV-dominated phase (small $R_A$, large $H$, where the entropy correction is subdominant) and an IR-dominated phase (large $R_A$, small $H$, where the entropy correction is dominant), rather than an artifact of a particular choice of equation of state. We do realize, however, that without a microscopic statistical mechanics for the degrees of freedom of the horizon, this interpretation has a partly phenomenological character, and have weakened the language in the conclusions accordingly.


\section{Classical stability}\label{sec6}

To better understand the role played by the parameters $c$, $K$ and $\beta$, we here analyze the minimum criterion to have a stable model. This is based on the sound speed of the content of the matter distribution. So, as usual, the sound speed is defined by  
\begin{equation}
v^{2}_{s}=\frac{dp_{de}}{\rho _{de}}=\frac{d\left( \omega _{de}d\rho _{de}\right) 
}{d\rho _{de}}=\omega _{de}+\rho _{de}\frac{d\omega _{de}}{d\rho _{de}}. 
\label{VI.1}
\end{equation}
Now, taking into account $\rho_{\text{de}}$ given by the expression \eqref{eq7} and the conservation Eq. \eqref{conservation}, one obtains

\begin{equation}
1+\omega _{de}=\frac{2}{3}\left( 1+q\right) \left( \frac{%
3c^{2}H^{2}-K^{2}/H^{2}}{\rho _{de}}\right).  \label{VI.3}
\end{equation}
On the other hand, by combining the Friedmann equations, the following result is obtained
\begin{equation}
1+q=\frac{3}{2}\left( 1+\frac{\omega _{de}\rho _{de}}{\rho +\rho _{de}}%
\right).  \label{VI.4}
\end{equation}
Therefore, replacing \eqref{VI.4} into  \eqref{VI.3}, one has

\begin{equation}
1+\omega _{de}=\left( 1+\frac{\omega _{de}\rho _{de}}{\rho +\rho _{de}}%
\right) \left( \frac{3c^{2}H^{2}-K^{2}/H^{2}}{\rho _{de}}\right) , 
\label{VI.5}
\end{equation}
or in a more compact fashion 
\begin{equation}
\omega _{de}=-\left( \frac{1+r}{1+r-A}\right) \left( 1-A\right),  \label{VI.6}
\end{equation}
where $A$ is being defined by 
\begin{equation}
A=\frac{3c^{2}H^{2}-K^{2}/H^{2}}{\rho _{de}}=\frac{3c^{2}H^{2}-K^{2}/H^{2}}{%
3c^{2}H^{2}+K^{2}/H^{2}}.  \label{VI.7}
\end{equation}
In general the expression \eqref{VI.7} can be recast as

\begin{equation}
A=1-\frac{6K^{2}}{\left( 1+r\right) \rho _{de}^{2}},  \label{VI.81}
\end{equation}
thus, the EoS parameter for the dark \eqref{VI.6} sector is given by

\begin{equation}
\omega _{de}\left( \rho _{de}\right) =-6K^{2}\left[ \frac{1+r}{r\left(
1+r\right) \rho _{de}^{2}+6K^{2}}\right].  \label{VI.10}
\end{equation}
Then
\begin{equation}
\frac{d\omega _{de}}{d\rho _{de}}=\omega _{de}\left( \frac{r}{1+r}\right)
\left( \frac{\omega _{de}}{6}\frac{\rho _{de}}{K^{2}}-\frac{1}{\rho _{de}}%
\right).  \label{VI.11}
\end{equation}
Returning to Eq. \eqref{VI.1} one gets

\begin{equation}
v^{2} =\frac{1}{6}\left( \frac{\left\vert \omega _{de}\right\vert }{K}%
\right) ^{2}\left( \rho _{de}^{2}-\frac{6K^{2}}{\left\vert \omega
_{de}\right\vert r}\right) \frac{r}{1+r}.  \label{VI.12} 
\end{equation}
Therefore, stability demands 
\begin{equation}
v^{2}_{s} >0\Longrightarrow \rho _{de}>\sqrt{\frac{6}{%
\left\vert \omega _{de}\right\vert r}}K.  \label{VI.13}
\end{equation}

Equation~\eqref{VI.13} provides a lower bound on the effective dark-energy density required to maintain classical stability whenever $v^{2}_{s}$ is interpreted as a squared effective sound speed. For further details, see Appendices \ref{app:A} and \ref{app:B}.

\section{Modified Friedmann dynamics and observational constraints}\label{sec7}
In this section we derive the background cosmological dynamics induced by the Kaniadakis holographic framework, focusing on the explicit form of the Hubble expansion rate \(H(z)\), which will later be confronted with observational data through a statistical analysis. We consider a spatially flat FLRW universe filled with a pressureless matter component and an effective dark-energy sector originating from the Kaniadakis entropy of the apparent horizon. The modified Friedmann equation can be written in the standard form
\begin{equation}
3H^{2} = \rho_{m} + \rho_{\text{de}},
\label{eq:F1-standard-form}
\end{equation}
where \(\rho_m\) denotes the matter energy density and \(\rho_\text{de}\) represents the effective Kaniadakis holographic contribution. In the following, we present two particular cases of \(\rho_\text{de}\) and the corresponding expressions for the Hubble function. For further details, see Appendix \ref{app:C}.
The first model is defined by expression \eqref{eq7} (Model 1 from now on)
and admits an analytic solution for the expansion rate, which can be expressed as
\begin{equation}
\dfrac{H^{2}(z)}{H^2_0} =
\frac{
\Omega_{m0}\,(1+z)^{3}
+\sqrt{\Omega_{m0}^{2}\,(1+z)^{6}
+\dfrac{2}{3}\,( 1 - c^{2})\dfrac{K^2}{H_0^4}\,}
}
{2\,(1-c^{2})}, \label{eq_model1}
\end{equation}
where, $K^{2} = 6 H_{0}^{4}(1 - c^{2} - \Omega_{m0})$.

For the model given by Eq.~ \eqref{eq45} (Model 2 from now on), 
the presence of a term proportional to \(\dot{H}\) leads to a first-order differential equation for the Hubble parameter. Rewriting the dynamics in terms of the redshift variable, one obtains
\begin{equation}
\frac{dH}{dz}
=
\frac{-3(1-c^{2})H^{2} + 3H_0^2\Omega_{m0}(1+z)^{3} + \dfrac{K^{2}}{H^{2}}}
{3\beta H(1+z)} . \label{eq_model2}
\end{equation}
In this case, the expansion history \(H(z)\) must be obtained numerically and will be employed in the observational analysis presented in the next section.

\section{Cosmological Perturbations and Structure Growth}
\label{sec8}
Having established the background dynamics and thermodynamic structure of the Kaniadakis HDE model, we now analyze the perturbative sector. This approach both checks the classical stability criterion from Sec.~\ref{sec6} and yields predictions for the growth rate of large-scale structure, offering a way to break the background degeneracy between $\Omega_{m0}$ and $c^2$ identified in Sec.~\ref{sec8}.
In order to study the perturbation equations in the Newtonian gauge, we work in the longitudinal gauge, where the
perturbed FLRW metric reads
\begin{equation}
ds^2 = -(1+2\Phi)\,dt^2 + a^2(t)(1-2\Phi)\,\delta_{ij}\,dx^i dx^j,
\label{eq:Newtonian_gauge}
\end{equation}
where $\Phi$ is the gauge-invariant Bardeen potential. For a
pressureless matter component, the linearized continuity and
Euler equations in Fourier space is given by
\begin{equation}
\ddot{\delta}_m + 2H\dot{\delta}_m
    = -\frac{k^2}{a^2}\Phi,
\label{eq:delta_m_general}
\end{equation}
while the perturbed Einstein equations yield the modified
Poisson equation
\begin{equation}
\frac{k^2}{a^2}\Phi
    = -\frac{1}{2}\left(\rho_m\delta_m + \rho_{de}\delta_m\right).
\label{eq:Poisson}
\end{equation}
A distinctive feature of the Kaniadakis holographic framework is
that the dark energy density (\ref{eq7}) is an explicit functional of the
Hubble parameter. Consequently, the DE perturbation $\delta\rho_{de}$ is entirely
determined by the local perturbation of the Hubble rate,
$\delta H = -\dot{\Phi} - H\Phi$, and possesses no independent
propagating degree of freedom. Explicitly,
\begin{equation}
\delta\rho_{de}
    = \left(6c^2H - \frac{2K^2}{H^3}\right)\delta H
    = \rho_{de}\,A(z)\,\frac{\delta H}{H},
\label{eq:delta_rhode}
\end{equation}
where we have defined the \emph{coupling parameter}
which is precisely the auxiliary quantity introduced in
Eq.~\eqref{VI.7} of the stability analysis. This identification is not
a coincidence: both the classical stability criterion and the
perturbative response of the dark energy is controlled by the
same competition between the ultraviolet term $3c^2H^2$ and the
infrared term $K^2/H^2$.
The parameter $A(z)$ satisfies $A \in (-1, +1)$ and admits a
transparent physical interpretation:
\begin{itemize}
\item $A < 0$ (IR-dominated regime, $K^2/H^2 > 3c^2H^2$):
  the dark energy perturbation opposes the gravitational
  collapse of matter, weakening the effective gravitational
  coupling.
\item $A > 0$ (UV-dominated regime, $K^2/H^2 < 3c^2H^2$):
  the dark energy perturbation reinforces gravitational
  collapse, enhancing structure growth.
\item $A = 0$: the UV and IR contributions balance exactly;
  the dark energy decouples from perturbations at this instant.
\end{itemize}

\begin{figure*}[t]
\centering
\includegraphics[width=0.95\textwidth]{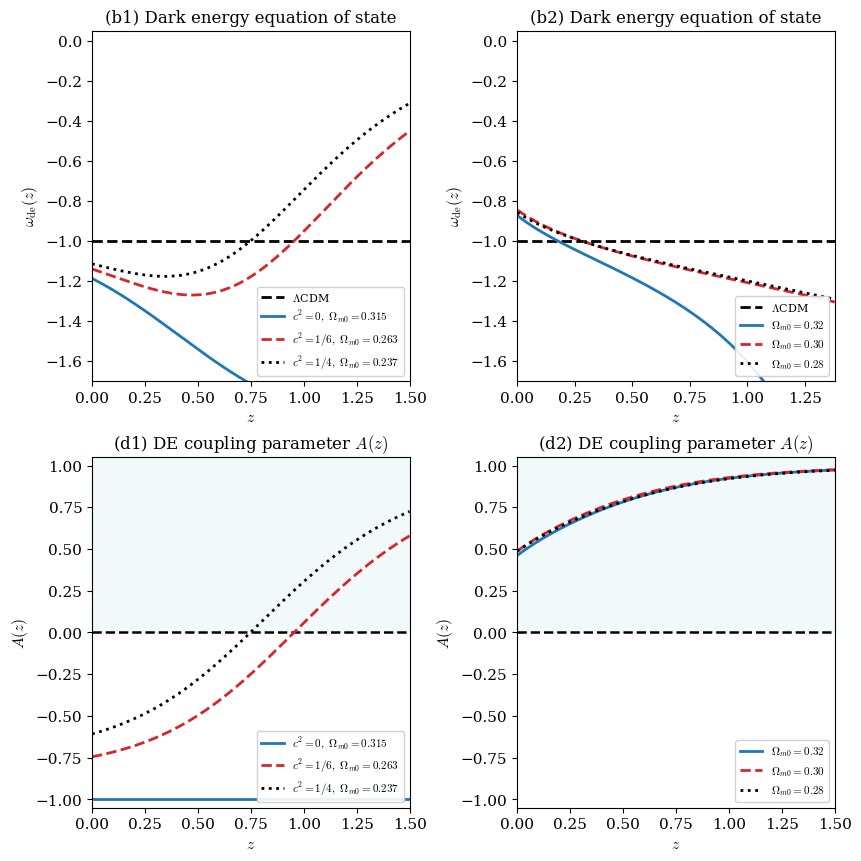}
\caption{{
Comparison of the dark energy equation of state and the geometric coupling parameter for the two Kaniadakis holographic dark energy models considered in this work.
Panels (b1) and (b2) display the evolution of the dark energy equation of state $\omega_{\rm de}(z)$ for Model~1 and Model~2, respectively, together with the $\Lambda$CDM reference $\omega=-1$ (black dashed line).
Panels (d1) and (d2) show the corresponding evolution of the effective coupling parameter
$A(z)=\frac{3c^2H^2-K^2/H^2}{3c^2H^2+K^2/H^2},$
which acts as a geometric order parameter measuring the competition between the ultraviolet holographic contribution and the infrared Kaniadakis correction.
For Model~1, $A(z)$ may cross zero depending on the value of $c^2$, signaling a transition between infrared-dominated ($A<0$) and ultraviolet-dominated ($A>0$) regimes and providing the dynamical counterpart of the geometric phase transition of the apparent horizon discussed in Sec.~V.
In contrast, Model~2 remains in the ultraviolet-dominated regime throughout the entire redshift interval, with $A(z)>0$ for all representative parameter choices.
The corresponding behavior of $\omega_{de}(z)$ reflects this distinction and illustrates how the thermodynamic structure of the horizon is directly encoded in the cosmological dynamics.
}}
\label{fig:bd_models_comparison}
\end{figure*}

It is useful to clarify the physical status of $A(z)$. On the perturbative level it manifests itself as a coupling parameter governing the response of dark-energy fluctuations to metric perturbations via $\delta\rho_{\rm de}=\rho_{\rm de}\,A(z)\,\delta H/H$. However, $A(z)$ has an inherent thermodynamic interpretation that is independent of the particular perturbation gauge or any particular choice of matter sector. Using (43) one can write 
\begin{equation} 
A \;=\; 1 - \frac{2\,K^2/H^2}{\rho_{de}} \;=\; \frac{\partial\ln\rho_{de}}{\partial\ln(3c^2H^2)} \bigg|_{K}, \label{eq:A_thermo} 
\end{equation} 
such that $A$ is the fractional weight of the Bekenstein-Hawking contribution $3c^2H^2$ with respect to the total dark energy density.
More fundamentally, from the use of the expansion~(3) we can recognize $3c^2H^2\propto S_{BH}$, $K^2/H^2\propto K^2S_{BH}^3/6$, so that $A$ measures the relative weight of the leading versus the subleading Kaniadakis entropic term. The zero crossing $A=0$, for $3c^2H^2=K^2/H^2$, defines the redshift $z_*$ when the two entropic contributions are in perfect balance. This is a property to be expected of the Kaniadakis entropy functional: it is the inflection point of $S_K(S_{BH})$ the variable $S_{BH}$ independent of the matter sector, the perturbation gauge, or the specific value of $c^2$. In this sense it $A(z)$ is not only a convenient parameterization but also reflects the entropic balance between the UV and IR sectors of the modified horizon thermodynamics. Its change of sign at $z=z_*$ is the perturbative analog of the thermodynamic phase transition found in Sec.~V: both are due to the same condition $3c^2H^2=K^2/H^2$, which is the crossing of the two scales generated by the Kaniadakis entropy, and both take place in the same redshift range $z\lesssim 1$, since this is exactly where the two entropic contributions have comparable values along the cosmological background.
In order to show  the connection between the background evolution and the geometric interpretation of the perturbative sector, in Fig.~\ref{fig:bd_models_comparison} we compare the evolution of the dark energy equation of state and the effective coupling parameter $A(z)$ for both models considered in this work. While Model~1 exhibits a possible sign reversal of $A(z)$, indicating a transition between infrared- and ultraviolet-dominated regimes, Model~2 remains entirely in the ultraviolet-dominated phase. This comparison highlights how the thermodynamic structure of the apparent horizon is reflected directly in the cosmological dynamics.
For the parameter combinations constrained in Sec.~\ref{sec8},
we find that $A(z)$ generically crosses zero during the
matter-dominated era, defining a \emph{perturbative transition
redshift} $z_*$ at which the growth dynamics changes
qualitatively. This transition is the perturbative analog of
the inverted thermodynamic phase transition identified in
Sec.~\ref{sec5}, and both originate from the same
UV/IR competition encoded in the Kaniadakis deformation.
Substituting Eq.~\eqref{eq:delta_rhode} into the Poisson
equation~\eqref{eq:Poisson} and using the quasi-static
approximation $\dot{\Phi} \ll H\Phi$ (valid for $k \gg aH$),
the system closes and yields a scale-dependent effective
gravitational coupling:
\begin{equation}
\frac{G_{\rm eff}}{G}
    \equiv \mu(k,z)
    = 1 + \frac{2}{3}\,\frac{\Omega_{de}(z)}{\Omega_m(z)}\,A(z)
       \frac{(aH/k)^2}{1+(aH/k)^2}.
\label{eq:Geff}
\end{equation}
On the sub-horizon scales relevant for RSD observations, $k \gg aH$ and therefore
\begin{equation}
\mu \to 1 \quad \text{(GR recovered)},
\label{eq:subhorizon}
\end{equation}
which confirms that the sub-horizon stability criterion $v_s^2 > 0$ derived in Eq.~\eqref{VI.13} is indeed sufficient for perturbative stability: in this limit general relativity is recovered and the Jeans instability is controlled solely by $v_s^2$.
In the sub-horizon limit, combining
Eqs.~\eqref{eq:delta_m_general} and \eqref{eq:Poisson} with
$\mu \to 1$ gives the standard growth equation with the
Kaniadakis background:
\begin{equation}
\delta_m'' + \left[\frac{3}{1+z} - \frac{H'(z)}{H(z)}\right]\delta_m'
    - \frac{3\,\Omega_m(z)}{2(1+z)^2}\,\mu(z)\,\delta_m = 0,
\label{eq:growth_ode}
\end{equation}
where primes denote derivatives with respect to $z$, and
$H(z)$ is given by Eq.~\eqref{eq_model1}. The key modification with respect
to $\Lambda$CDM enters through two channels: (i) the altered
expansion history $H(z)$, which changes the friction term
$H'(z)/H(z)$, and (ii) the modified matter fraction
$\Omega_m(z) = 1 - \Omega_{de}(z)$ with
$\Omega_{de}(z) = c^2 + K^2/(3H^4)$.
The growth rate and the observable $f\sigma_8$ are defined as
\begin{equation}
f(z) \equiv \frac{d\ln\delta_m}{d\ln a}
           = -\frac{1+z}{\delta_m}\,\frac{d\delta_m}{dz},\qquad
f\sigma_8(z) = f(z)\cdot\sigma_8(z),
\label{eq:fsigma8_def}
\end{equation}
where $\sigma_8(z) = \sigma_{8,0}\,\delta_m(z)/\delta_m(0)$.

{For representative values of $c^2$ within the thermodynamically 
allowed range (Sec.~\ref{sec5}), the coupling parameter $A(z)$ 
generically crosses zero during the matter-dominated era at a 
\emph{perturbative transition redshift} $z_*$: numerically, 
$z_* \approx 0.7$ for $c^2 = 1/6$ and $z_* \approx 0.4$ for 
$c^2 = 1/4$, while for $c^2 = 0$ the infrared term dominates 
at all redshifts and $A(z) = -1$ throughout. The sign change 
of $A(z)$ signals a qualitative transition in the perturbative 
sector: at $z > z_*$ the dark energy perturbation reinforces 
gravitational collapse, while at $z < z_*$ it opposes it. 
This transition is the perturbative counterpart of the inverted 
thermodynamic phase transition identified in Sec.~\ref{sec5}, 
both originating from the same UV/IR competition encoded in 
the Kaniadakis deformation and occurring in the same redshift 
range. The specific values of $c^2$ that are compatible with 
the data are determined in the next section.}
\section{Statistical analysis: $\chi^2$ estimators}
\label{sec9}
In order to constrain the free parameters of the model, we perform a joint likelihood analysis using three independent late-time cosmological probes: Cosmic Chronometers (CC), Type Ia Supernovae (SNe Ia), and Baryon Acoustic Oscillations (BAO) from the DESI DR2 survey. These datasets provide complementary information on the cosmic expansion history and allow us to assess parameter degeneracies in a controlled manner.
Assuming statistical independence among the different probes, the total chi-square function is defined as
\begin{equation}
\chi^2_{\rm tot} = \chi^2_{\rm CC} + \chi^2_{\rm SN} + \chi^2_{\rm DESI}.
\end{equation}
In the following, we describe the construction of each individual contribution.
\subsection*{Cosmic Chronometers (CC)}
Cosmic chronometers (CC) provide a method to determine the Hubble parameter as a function of redshift in a largely model-independent way. This approach is based on the assumption that the expansion of the Universe causes the ages of passively evolving galaxies to vary with redshift, allowing the expansion rate to be inferred without relying on any cosmological distance ladder.
We employ a compilation of $N_{\rm CC}$ measurements $\{z_i, H_{\rm obs}(z_i), \sigma_{H_i}\}$ from the literature \cite{Moresco:2020fbm,cao2018cosmological,farooq2013hubble}.
The corresponding chi-square estimator is given by
\begin{equation}
\chi^2_{\rm CC} =
\sum_{i=1}^{N_{\rm CC}}
\frac{\left[H_{\rm th}(z_i;\mathbf{p}) - H_{\rm obs}(z_i)\right]^2}{\sigma_{H_i}^2},
\end{equation}
where $H_{\rm th}(z_i;\mathbf{p})$ denotes the theoretical prediction of the Hubble parameter for a given set of model parameters $\mathbf{p}$.
\subsection*{Type Ia Supernovae (SNe Ia)}
Type Ia supernovae provide precise measurements of relative cosmological distances through the distance modulus,
\begin{equation}
\mu_{\rm obs} = m_B^{\rm corr} - M,
\end{equation}
where $m_B^{\rm corr}$ is the corrected apparent magnitude and $M$ denotes the absolute magnitude of the standard candle.
The theoretical distance modulus is computed as
\begin{equation}
\mu_{\rm th}(z;\mathbf{p}) = 5 \log_{10}\!\left[\frac{D_L(z;\mathbf{p})}{\mathrm{Mpc}}\right] + 25,
\end{equation}
with the luminosity distance given by
\begin{equation}
D_L(z;\mathbf{p}) = (1+z)\int_0^z \frac{c\,dz'}{H(z';\mathbf{p})}.
\end{equation}
We use the PantheonPlus compilation \cite{Brownsberger:2021uue,Brout:2022vxf,Scolnic:2021amr}\footnote{Available at \url{https://github.com/PantheonPlusSH0ES}.}, which contains 1657 SNe Ia with $z>0.01$.
The associated chi-square function is defined as
\begin{equation}
\chi^2_{\rm SN} =
(\boldsymbol{\mu}_{\rm obs} - \boldsymbol{\mu}_{\rm th})^{T}
\,\mathbf{C}^{-1}\,
(\boldsymbol{\mu}_{\rm obs} - \boldsymbol{\mu}_{\rm th}),
\end{equation}
where $\mathbf{C}$ is the full covariance matrix including both statistical and systematic uncertainties.
It is important to note that SNe Ia constrain only relative distances and are therefore insensitive to the absolute expansion scale. As a consequence, the parameters $M$ and $H_0$ appear in a degenerate combination, which cannot be broken by supernova data alone. In our analysis, this degeneracy is lifted only when SNe Ia are combined with probes that provide absolute measurements of the expansion rate.
\subsection*{Baryon Acoustic Oscillations (DESI)}
Baryon acoustic oscillations constitute a robust standard ruler for cosmological distance measurements. We employ the latest BAO data from the DESI DR2 release \cite{DESI:2025zgx,DESI:2025zpo}\footnote{Available at \url{https://github.com/CobayaSampler/bao_data/}.}, which provide separate measurements of the transverse comoving distance and the Hubble distance.
The observational data vector is defined as
\begin{equation}
\mathbf{X}_{\rm obs} =
\left\{
\frac{D_M(z)}{r_d},\,
\frac{D_H(z)}{r_d},\,
\frac{D_V(z)}{r_d}
\right\}_{\rm obs},
\end{equation}
where
$D_M(z) = \int_0^{z} \frac{c}{H(z')}\,dz'$ is the transverse comoving distance,
$D_H(z)=c/H(z)$ is the Hubble distance,
and $D_V(z) = \left[ D_M^2(z)\,\frac{c\,z}{H(z)} \right]^{1/3}.
$ denotes the volume-averaged distance.
The theoretical counterpart $\mathbf{X}_{\rm th}$ is constructed analogously.
The BAO chi-square is given by
\begin{equation}
\chi^2_{\rm DESI} =
(\mathbf{X}_{\rm obs} - \mathbf{X}_{\rm th})^{T}
\,\mathbf{C}^{-1}\,
(\mathbf{X}_{\rm obs} - \mathbf{X}_{\rm th}),
\end{equation}
where $\mathbf{C}$ is the covariance matrix provided by the DESI collaboration. {}
Since the model considered in this work is defined only at late times and does not modify the early-Universe physics that determines the sound horizon, we fix the sound horizon at the drag epoch to its $\Lambda$CDM value,
$r_d = 147.09\,\mathrm{Mpc}$ \cite{Planck:2018vyg}.

\subsection*{Redshift-space distortions ($f\sigma_8$)}
The growth of cosmic structure is probed through the compilation of redshift-space distortion (RSD) measurements $\mathbf{d} = \{f\sigma_8(z_i)\}_{i=1}^N$ with covariance matrix $\mathbf{C}$~\cite{Kazantzidis:2018rnb}. For the parameter vector $\mathbf{p}$, the theoretical counterpart $m_i(\mathbf{p}) = f\sigma_8(z_i; \mathbf{p})$ follows from Eq.~\eqref{eq:fsigma8_def}, leading to
\begin{equation}
\chi^2_\mathrm{RSD} = (\mathbf{d} - \mathbf{m})^T\,\mathbf{C}^{-1}\,(\mathbf{d} - \mathbf{m}).
\end{equation}

\subsection{Model 1}
\begin{figure}[!t]
    \centering
    \includegraphics[width=0.45\textwidth]{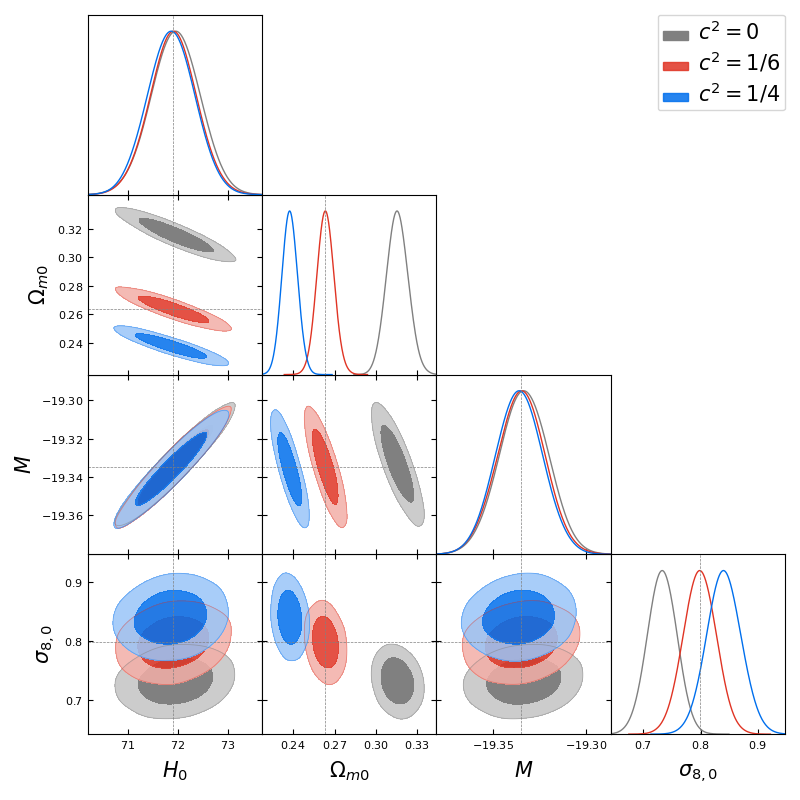}
    \caption{
    Triangle plot showing the joint constraints from CC + PantheonPlus + DESI for three fixed values of the parameter $c^2$, for model 1, Eq. \eqref{eq_model1}.
    While the inferred value of the present-day matter density $\Omega_{m0}$ shifts systematically as $c$ varies, the posterior distributions of the Hubble constant $H_0$ and the supernova absolute magnitude $M$ remain unchanged.
    This behavior reflects the exact degeneracy between $\Omega_{m0}$ and $c$ at the background level, indicating that late-time expansion data are insensitive to $c$.
    }
    \label{fig:triangle1_fixed_c}
\end{figure}

Figure~\ref{fig:triangle1_fixed_c} and Table~\ref{tab:fixed_c_results} show the constraints obtained from CC + PantheonPlus + DESI RD2 + RSD for three fixed values of the parameter $c^2$. The marginal posteriors of $H_0$ and the supernova absolute magnitude $M$ are essentially identical across the three scenarios, and the two-dimensional $M$--$H_0$ contours overlap almost completely, forming a single elongated band along the standard SN anticorrelation. In contrast, the marginal posteriors of $\Omega_{m0}$ are well separated, with three disjoint peaks at $\Omega_{m0} \simeq 0.32, 0.26, 0.21$ for $c^2 = 0, 1/6, 1/3$ respectively. This behaviour reflects the exact background degeneracy between $\Omega_{m0}$ and $c$: variations in $c$ are compensated by a rescaling of $\Omega_{m0}$, leaving the expansion history and the goodness of fit unchanged, as confirmed by the nearly identical $\chi^2_\nu \approx 0.935$ across the three columns.

The inclusion of RSD measurements through $\sigma_{8,0}$ adds further information. The inferred amplitude shifts from $\sigma_{8,0} = 0.733 \pm 0.023$ for $c^2 = 0$ to $\sigma_{8,0} = 0.891 \pm 0.028$ for $c^2 = 1/3$, while $H_0$ and $M$ remain fixed. This sensitivity reflects the fact that the growth of structure depends on the dark energy fraction $\Omega_{de}(z)$,which is modified when $c$ changes. The triangle plot in Fig.~\ref{fig:triangle1_fixed_c} shows that the marginal posteriors of $\sigma_{8,0}$ are clearly displaced, with overlap only in the distribution tails, and the two-dimensional contours in the $\sigma_{8,0}$--$\Omega_{m0}$ plane separate into three well-resolved regions. In the planes $\sigma_{8,0}$--$H_0$ and $\sigma_{8,0}$--$M$, however, the three contours remain stacked vertically and overlap in the $\sigma_{8,0}$ direction, indicating that RSD does not constrain $c^2$ uniquely: the data are equally well reproduced by combinations in which a higher $c^2$ is compensated by a lower $\Omega_{m0}$ and a higher $\sigma_{8,0}$.

\begin{figure}[!t]
    \centering
    \includegraphics[width=0.45\textwidth]{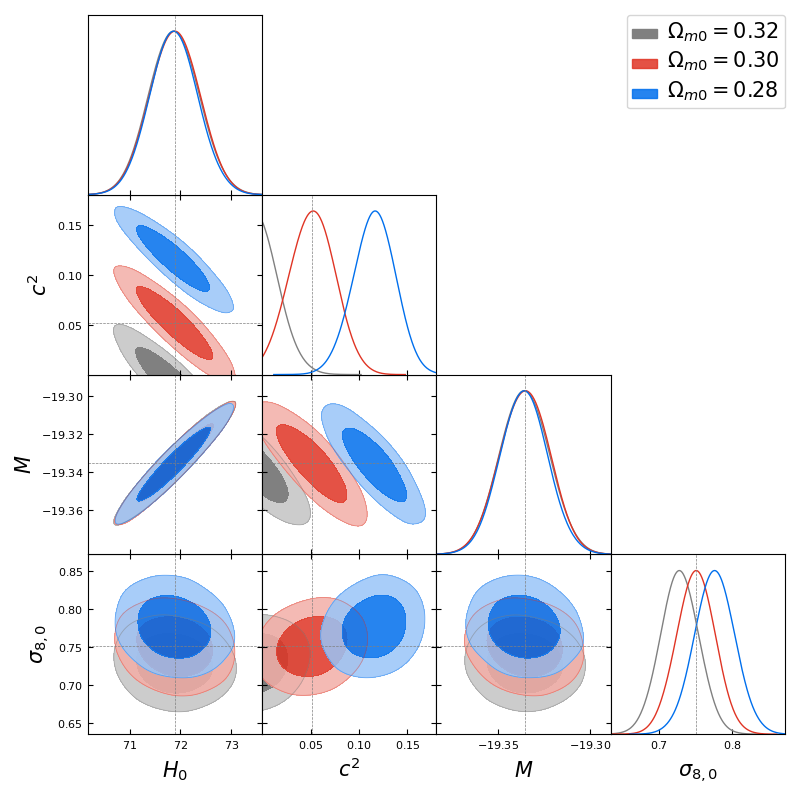}
    \caption{
    Triangle plot showing the joint constraints from CC + PantheonPlus + DESI for three fixed values of the present-day matter density $\Omega_{m0}$, for model 1, Eq. \eqref{eq_model1}.
    In this case, apparently well-defined posterior distributions for the parameter $c$ emerge.
    However, these constraints are entirely driven by the imposed value of $\Omega_{m0}$ and do not originate from the data themselves.
    This behaviour reflects the exact degeneracy between $\Omega_{m0}$ and $c$ at the background level: once $\Omega_{m0}$ is fixed, the value of $c$ is algebraically determined, while the expansion history and the quality of the fit remain unchanged.
    }
    \label{fig:triangle1_fixed_om0}
\end{figure}

Figure~\ref{fig:triangle1_fixed_om0} and Table~\ref{tab:fixed_om0_results} illustrate the complementary situation in which the present-day matter density $\Omega_{m0}$ is fixed to different values, while the parameter $c$ is treated as a free variable. The contrast with Fig.~\ref{fig:triangle1_fixed_c} is striking: in this case, the marginal posteriors of $\sigma_{8,0}$ are no longer separated. The three distributions, centred at $0.728 \pm 0.023$, $0.750 \pm 0.024$, and $0.776 \pm 0.024$, overlap heavily, and the corresponding two-dimensional contours in the $\sigma_{8,0}$--$H_0$ and $\sigma_{8,0}$--$M$ planes are nearly indistinguishable. The posteriors of $c^2$ themselves, while shifted across the three scenarios, also show non-negligible overlap, with the $\Omega_{m0} = 0.32$ case piling against the prior boundary at $c^2 = 0$. The apparent constraints on $c$ in this configuration are therefore driven by the imposed value of $\Omega_{m0}$ rather than by the data: the narrow range of $\Omega_{m0}$ explored here probes a trajectory close to the degeneracy direction selected by the background data, along which the growth predictions are nearly indistinguishable.

Taken together, Figures~\ref{fig:triangle1_fixed_c} and~\ref{fig:triangle1_fixed_om0} show that the combination CC + SN + BAO + RSD constrains the model along a degenerate direction in the joint $(c, \Omega_{m0}, \sigma_{8,0})$ space rather than in any of these parameters individually. Fixing $c$ at different values selects three well-separated points along this direction (Fig.~\ref{fig:triangle1_fixed_c}), whereas fixing $\Omega_{m0}$ within its observationally allowed range explores positions close to the degeneracy itself, producing strongly overlapping posteriors (Fig.~\ref{fig:triangle1_fixed_om0}). The reduced chi-squared remains nearly constant in both analyses ($\chi^2_\nu \approx 0.935$ in all cases), confirming that no scenario is statistically preferred. Breaking this three-parameter degeneracy would require complementary information from high-redshift probes such as CMB anisotropies, or independent priors on $\sigma_{8,0}$.
Our results can be compared with those of \cite{Hernandez-Almada:2021aiw}, which performed an MCMC analysis of Kaniadakis HDE using CC, SNe~Ia and BAO data (without DESI DR2 or RSD). The work found $K\approx 0$ at $1\sigma$ level, consistent with our Tables I-II, where the derived quantity $K^2/H_0^4$ goes to $0$ in limit $c^2\to 1-\Omega_{m0}$. The main difference is that \cite{Hernandez-Almada:2021aiw} considered the future event horizon as the infrared cutoff, while we consider the Hubble horizon, $L=H^{-1}$ which produces the explicit $H^{-2}$ infrared correction $\rho_{\rm de}$ and the thermodynamic phenomenology analyzed in Sections~III-V.

\begin{table}[!t]
\centering
\caption{Constraints from CC + PantheonPlus + DESI for different fixed values of $c^2$.
We report mean values with $68\%$ confidence levels, best-fit values for the free parameters,
and the corresponding best-fit value of $K/H_0^4$.
}
\label{tab:fixed_c_results}
\begin{tabular}{lccc}
\hline\hline
Parameter & $c^{2}=0$ & $c^{2}=1/6$ & $c^{2}=1/4$ \\
\hline
\multicolumn{4}{c}{\textit{Mean values and 68\% CL}} \\
\hline
$H_{0}$        & $71.95 \pm 0.44$     & $71.91 \pm 0.42$     & $71.87 \pm 0.43$ \\
$\Omega_{m0}$  & $0.316 \pm 0.007$    & $0.263 \pm 0.006$    & $0.238 \pm 0.005$ \\
$M$            & $-19.333 \pm 0.012$  & $-19.335 \pm 0.012$  & $-19.336 \pm 0.012$ \\
$\sigma_{8, 0}$   & $0.733 \pm 0.023$    & $0.799 \pm 0.026$    & $0.841 \pm 0.027$ \\
\hline
\multicolumn{4}{c}{\textit{Best-fit values}} \\
\hline
$H_{0}$        & $71.96$   & $71.95$   & $71.88$ \\
$\Omega_{m0}$  & $0.315$   & $0.263$   & $0.237$ \\
$M$            & $-19.333$ & $-19.334$ & $-19.335$ \\
$\sigma_{8, 0}$   & $0.733$   & $0.799$   & $0.840$ \\
\hline
\multicolumn{4}{c}{\textit{Derived quantity (best fit)}} \\
\hline
$K^{2}/H_{0}^{4}$ & $4.11$ & $3.42$ & $3.08$ \\
\hline
\multicolumn{4}{c}{\textit{Goodness of fit}} \\
\hline
$\chi^{2}_{\min}$ & $1588.24$ & $1585.88$ & $1584.93$ \\
$\chi^{2}_{\nu}$  & $0.936$   & $0.935$   & $0.934$ \\
\hline\hline
\end{tabular}
\end{table}






\begin{table}[t]
\centering
\caption{Constraints from CC + PantheonPlus + DESI for different fixed values of $\Omega_{m0}$.
We report mean values with $68\%$ confidence levels, best-fit values for the free parameters,
and the corresponding best-fit value of $K/H_0^4$.}
\label{tab:fixed_om0_results}
\begin{tabular}{lccc}
\hline\hline
Parameter & $\Omega_{m0}=0.32$ & $\Omega_{m0}=0.30$ & $\Omega_{m0}=0.28$ \\
\hline
\multicolumn{4}{c}{\textit{Mean values and 68\% CL}} \\
\hline
$H_{0}$       & $71.88 \pm 0.44$     & $71.89 \pm 0.44$     & $71.87 \pm 0.42$ \\
$c^{2}$       & $0.011 \pm 0.025$   & $0.051 \pm 0.022$    & $0.116 \pm 0.020$ \\
$M$           & $-19.335 \pm 0.012$  & $-19.335 \pm 0.012$  & $-19.336 \pm 0.012$ \\
$\sigma_{8, 0}$  & $0.728 \pm 0.023$    & $0.750 \pm 0.024$    & $0.776 \pm 0.024$ \\
\hline
\multicolumn{4}{c}{\textit{Best-fit values}} \\
\hline
$H_{0}$       & $71.88$   & $71.87$   & $71.87$ \\
$c^{2}$       & $0.012$  & $0.054$   & $0.116$ \\
$M$           & $-19.336$ & $-19.336$ & $-19.336$ \\
$\sigma_{8, 0}$  & $0.728$   & $0.751$   & $0.774$ \\
\hline
\multicolumn{4}{c}{\textit{Derived quantity (best fit)}} \\
\hline
$K^{2}/H_{0}^{4}$ & $4.01$ & $3.88$ & $3.62$ \\
\hline
\multicolumn{4}{c}{\textit{Goodness of fit}} \\
\hline
$\chi^{2}_{\min}$ & $1588.45$ & $1587.46$ & $1586.56$ \\
$\chi^{2}_{\nu}$  & $0.936$   & $0.935$   & $0.935$ \\
\hline\hline
\end{tabular}
\end{table}






\subsection{Model 2}
\begin{figure}[H]
    \centering
    \includegraphics[width=0.45\textwidth]{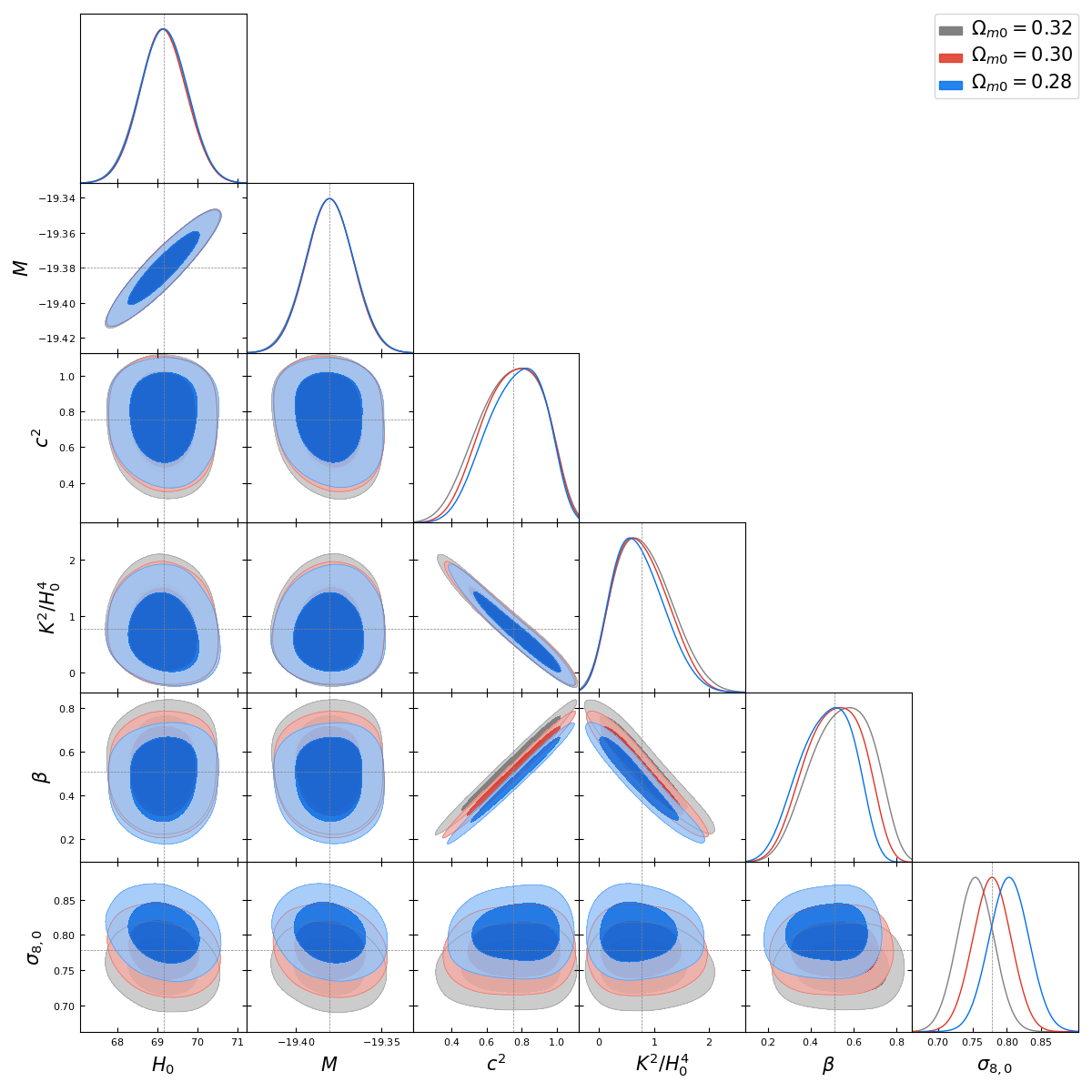}
    \caption{Posterior distributions and two-dimensional marginalized confidence regions at $68\%$ and $95\%$ CL for model 2, Eq. \eqref{eq_model2}, parameters 
    $(H_0, M, c^2, K^2/H_0^4, \beta)$ 
    obtained from the joint CC+PantheonPlus+DESI analysis.
    The different colors correspond to fixed values of the matter density parameter 
    $\Omega_{m0}=0.28$ (blue), $0.30$ (red), and $0.32$ (gray).
    While the constraints on $H_0$ and $M$ remain remarkably stable, the parameters 
    $c^2$, $K^2/H_0^4$, and $\beta$ exhibit significant degeneracies, indicating that late-time background observables alone are insufficient to break these correlations.
    }
    \label{fig:corner_5param_beta}
\end{figure}

Figure~\ref{fig:corner_5param_beta} and Table~\ref{tab:fixed_om0_beta_model} present the constraints obtained from the joint CC + PantheonPlus + DESI RD2 + RSD analysis for the extended model including the additional parameter $\beta$, for three fixed values of the present-day matter density $\Omega_{m0}$. The posterior distributions of the Hubble constant $H_0$ and the supernova absolute magnitude $M$ are essentially insensitive to $\Omega_{m0}$: the three marginals overlap to the point of being indistinguishable ($H_0 \simeq 69.15 \pm 0.52$, $M \simeq -19.380 \pm 0.012$), and the two-dimensional $M$--$H_0$ contour appears as a single anticorrelated ellipse common to the three cases. The inferred central values are systematically shifted with respect to Model~1, reflecting the impact of the additional parameter on the late-time expansion.

The marginal posteriors of $c^2$ and $K^2/H_0^4$ are nearly identical across the three values of $\Omega_{m0}$ and individually broad, with central values $c^2 = 0.74 \pm 0.18$ and $K^2/H_0^4 = 0.80 \pm 0.50$ at the $1\sigma$ level (Table~\ref{tab:fixed_om0_beta_model}), indicating that neither parameter is independently constrained by the data. Their joint posterior in the $c^2$--$K^2/H_0^4$ plane, however, traces a sharp anticorrelation, reflecting the fact that the late-time expansion fixes a particular combination of these parameters rather than each one separately. A second, even narrower degeneracy appears in the $\beta$--$c^2$ plane as an elongated banana of positive slope, with the three fixed-$\Omega_{m0}$ contours visibly displaced along it; a corresponding anticorrelation appears in the $\beta$--$K^2/H_0^4$ plane. The parameter $\beta$ itself is only mildly displaced across the three scenarios ($\beta = 0.55 \pm 0.14$, $0.51 \pm 0.13$, $0.51 \pm 0.13$ for $\Omega_{m0} = 0.32, 0.30, 0.28$), with marginals that overlap substantially. We note that the mean values for $\Omega_{m0} = 0.30$ and $\Omega_{m0} = 0.28$ are nearly indistinguishable across all model parameters, while $\Omega_{m0} = 0.32$ shows a small systematic shift, suggesting that the degeneracy direction in parameter space is locally aligned with the lower end of the $\Omega_{m0}$ interval explored here.

The inclusion of RSD measurements through $\sigma_{8,0}$ adds only marginal discriminating power in this model. The inferred amplitude shifts mildly across the three scenarios, $\sigma_{8,0} = 0.754 \pm 0.024$, $0.778 \pm 0.025$, and $0.778 \pm 0.025$ for $\Omega_{m0} = 0.32, 0.30, 0.28$ respectively, in the opposite direction to $\beta$. The corresponding marginal posteriors overlap significantly, and the two-dimensional contours in the $\sigma_{8,0}$--$c^2$, $\sigma_{8,0}$--$K^2/H_0^4$, and $\sigma_{8,0}$--$\beta$ planes show the three levels stacked vertically with strong overlap rather than forming well-resolved regions. The growth data are therefore absorbed by a coordinated shift of $\sigma_{8,0}$ along the multi-dimensional degeneracy, without isolating any of the additional parameters.

Taken together, these findings show that promoting the deformation parameter $\beta$ to a free quantity preserves the constraining power of the data while reshaping the structure of the degeneracies. Unlike Model~1, where fixing $\Omega_{m0}$ determined $c^2$ uniquely through an exact background degeneracy, the data now provide genuine constraints on each of the deformation parameters: the marginal posteriors of $c^2 = 0.74 \pm 0.18$ and $\beta = 0.55 \pm 0.14$ peak at well-defined non-zero values, while $K^2/H_0^4 = 0.80 \pm 0.50$ is more loosely constrained but still favours non-zero values. These constraints are essentially independent of the imposed value of $\Omega_{m0}$ within the explored range, as confirmed by the near-identity of the posteriors for $\Omega_{m0} = 0.30$ and $\Omega_{m0} = 0.28$. The remaining freedom resides in the internal correlations among the deformation parameters themselves: a sharp anticorrelation in the $c^2$--$K^2/H_0^4$ plane and a narrow positive correlation in the $\beta$--$c^2$ plane, along which different combinations yield comparable expansion and growth histories. The reduced chi-squared remains nearly constant across the three scenarios ($\chi^2_\nu \approx 0.896$, see Table~\ref{tab:fixed_om0_beta_model}), slightly improved with respect to Model~1 owing to the additional flexibility introduced by $\beta$. Complementary high-redshift information, such as CMB anisotropies or independent priors on $\sigma_{8,0}$, would help tighten the constraints along the degeneracy directions identified here and further test the model.
\begin{table}[H]
\centering
\caption{Constraints from CC + PantheonPlus + DESI for the second model including the parameter $\beta$, 
for different fixed values of the present-day matter density $\Omega_{m0}$. 
We report mean values with $68\%$ confidence levels and the corresponding best-fit values.}
\label{tab:fixed_om0_beta_model}
\begin{tabular}{lccc}
\hline\hline
Parameter & $\Omega_{m0}=0.32$ & $\Omega_{m0}=0.30$ & $\Omega_{m0}=0.28$ \\
\hline
\multicolumn{4}{c}{\textit{Mean values and 68\% CL}} \\
\hline
$H_{0}$        & $69.14 \pm 0.53$     & $69.15 \pm 0.52$     & $69.15 \pm 0.52$ \\
$M$            & $-19.380 \pm 0.012$  & $-19.380 \pm 0.012$  & $-19.380 \pm 0.012$ \\
$c^{2}$        & $0.740 \pm 0.186$    & $0.753 \pm 0.175$    & $0.753 \pm 0.175$ \\
$K^{2}/H_{0}^{4}$ & $0.82 \pm 0.51$   & $0.78 \pm 0.48$     & $0.78 \pm 0.48$ \\
$\beta$        & $0.546 \pm 0.142$    & $0.510 \pm 0.134$    & $0.510 \pm 0.134$ \\
$\sigma_{8, 0}$   & $0.754 \pm 0.024$    & $0.778 \pm 0.025$    & $0.778 \pm 0.025$ \\
\hline
\multicolumn{4}{c}{\textit{Best-fit values}} \\
\hline
$H_{0}$        & $69.14$   & $69.06$   & $69.13$ \\
$M$            & $-19.381$ & $-19.383$ & $-19.380$ \\
$c^{2}$        & $0.853$   & $0.818$   & $0.794$ \\
$K^{2}/H_{0}^{4}$ & $0.507$ & $0.592$ & $0.641$ \\
$\beta$        & $0.633$   & $0.555$   & $0.492$ \\
$\sigma_{8, 0}$   & $0.756$   & $0.778$   & $0.809$ \\
\hline
\multicolumn{4}{c}{\textit{Goodness of fit}} \\
\hline
$\chi^{2}_{\min}$ & $1522.00$ & $1521.01$ & $1520.14$ \\
$\chi^{2}_{\nu}$  & $0.897$   & $0.896$   & $0.896$ \\
\hline\hline
\end{tabular}
\end{table}

{
\subsection{Dark energy equation of state and consistency 
with DESI DR2}
It is worth emphasizing that the dynamical behavior of 
$w_{\rm DE}(z)$ exhibited by both models, as shown in 
Fig.~\ref{fig:bd_models_comparison}, arises naturally 
from the modified statistical framework, without invoking 
any phenomenological parameterization such as the CPL 
ansatz~\cite{Chevallier:2000qy,Linder:2002et}, $w(a) = w_0 + 
w_a(1-a)$. The best-fit values obtained from the joint 
CC + PantheonPlus + DESI DR2 analysis---summarized in 
Tables~\ref{tab:fixed_c_results} and~\ref{tab:fixed_om0_results}---yield 
an effective dark energy equation of state that deviates 
from $w=-1$ and exhibits a non-trivial redshift evolution, 
including phantom-like dynamics and crossing of the $w=-1$ 
line. We stress that this 
behavior is to be understood as \textit{effective}: the 
Kaniadakis-modified Friedmann equations can be recast in a 
form that mimics $\Lambda$CDM supplemented by a dynamical dark 
energy component, so that no fundamental phantom field with 
negative kinetic energy is introduced. 
Notably, the DESI DR2 extended dark energy analysis finds 
a robust preference for models featuring phantom crossing, 
a trend confirmed by non-parametric reconstructions of 
$w(z)$~\cite{DESI:2025fii,DESI:2025wyn} that are themselves 
independent of the CPL framework~\cite{DESI:2025zgx}. 
The fact that these models reproduce these qualitative trends 
without any \textit{a priori} assumption on the time 
evolution of dark energy constitutes a non-trivial result: 
the Kaniadakis framework is not fitted to mimic dynamical 
dark energy, yet it naturally accounts for it, strengthening 
the case for Kaniadakis statistics as a theoretically 
motivated alternative to purely phenomenological descriptions 
of late-time cosmic acceleration.
}

{\subsection{Bayesian model comparison}
\label{sec:bayesian_comparison}
To assess whether the improvement in the fit provided by the
Kaniadakis models is sufficient to compensate for their additional
degrees of freedom, we perform a Bayesian model comparison
\cite{Trotta:2008qt,Liddle:2007fy} against three reference cosmologies:
the standard $\Lambda$CDM model, the CPL parametrization, and the
conventional holographic dark energy (HDE) model.
All models
are confronted with the same combined data set,
\begin{equation}
{\cal D}={\rm CC+PantheonPlus+DESI~DR2+RSD},
\end{equation}
using identical treatments of the common nuisance parameters and
observational covariance matrices.
For a model ${\cal M}$ with parameters $\boldsymbol{\theta}$, the
Bayesian evidence is defined as
\begin{equation}
Z_{\cal M}
=
\int
{\cal L}({\cal D}|\boldsymbol{\theta},{\cal M})
\pi(\boldsymbol{\theta}|{\cal M})
\,d\boldsymbol{\theta},
\end{equation}
where ${\cal L}$ and $\pi$ denote the likelihood and prior,
respectively. We report the logarithmic evidence relative to
$\Lambda$CDM,
\begin{equation}
\Delta\ln Z_{\cal M}
=
\ln Z_{\cal M}
-
\ln Z_{\Lambda{\rm CDM}}.
\end{equation}
For interpretation, we regard $|\Delta\ln Z|<1$ as inconclusive,
$1<|\Delta\ln Z|<2.5$ as weak evidence,
$2.5<|\Delta\ln Z|<5$ as moderate evidence, and
$|\Delta\ln Z|>5$ as strong evidence against the model with the
smaller evidence. As complementary diagnostics, we also compute the Akaike and Bayesian
information criteria,
\begin{equation}
{\rm AIC}=\chi^2_{\min}+2k,
\qquad
{\rm BIC}=\chi^2_{\min}+k\ln N,
\end{equation}
where $k$ is the number of free parameters and $N=1720$ is the total
number of data points.
The Bayesian evidences were computed with the \texttt{dynesty}
nested-sampling algorithm, using 1000 live points, the multi-ellipsoidal
bounding scheme, random-walk sampling, and the stopping criterion
$\Delta \ln Z < 0.1$. The common
parameters were assigned the same priors in all models,
\begin{equation}
50<H_0<75,\qquad
0.10<\Omega_{m0}<0.50,
\end{equation}
\begin{equation}
-21<M<-17,\qquad
0.4<\sigma_{8,0}<1.5.
\end{equation}
For the CPL model we adopted
$-2<w_0<0$ and $-3<w_a<3$, while for standard HDE we used
$0.01<c_{\rm HDE}<2$. For the Kaniadakis parameters we adopted
$0<c^2<1$, $0<\kappa<10$, and $0.01<\beta<2$, whenever the
corresponding parameters are independent. Here
$\kappa\equiv K^2/H_0^4$.
In Model 1, $\kappa$ is not independent but is fixed by the
normalization condition
\begin{equation}
\kappa=6(1-c^2-\Omega_{m0}).
\end{equation}
Consequently, the prior was normalized directly over the physically
admissible region
$0.10<\Omega_{m0}<0.50$ and
$0<c^2<1-\Omega_{m0}$.
No additional restriction associated with the existence of the
thermodynamic critical point was imposed in the Bayesian analysis,
since such a condition would select only a subset of the cosmological
parameter space and would modify the prior volume entering the
evidence.
The resulting model comparison is summarized in
Table~\ref{tab:bayesian_comparison}. The reference $\Lambda$CDM model
gives
$\chi^2_{\min}=1520.569$ and
$\ln Z=-776.450\pm0.151$.
Standard HDE provides an essentially comparable maximum-likelihood
fit, with $\Delta\chi^2=-0.438$, but its additional parameter leads to
$\Delta\ln Z=-2.381\pm0.220$. The CPL parametrization improves the
minimum chi-square by only $\Delta\chi^2=-1.189$ while introducing two
additional parameters, resulting in
$\Delta\ln Z=-3.710\pm0.224$.
Model 2 exhibits a qualitatively different behavior. It reaches the
lowest minimum chi-square among the five cosmologies,
\begin{equation}
\chi^2_{\min}=1515.548,
\qquad
\Delta\chi^2=-5.021,
\end{equation}
showing that the additional Kaniadakis sector can improve the
maximum-likelihood description of the combined data. Nevertheless,
this improvement is not sufficient to compensate for the larger
parameter volume. We obtain
\begin{equation}
\Delta\ln Z_{\rm KHDE2}
=
-4.357\pm0.230.
\end{equation}
Thus, current data favor $\Lambda$CDM in the Bayesian comparison even
though Model 2 provides a better best fit. This distinction between an
improved maximum-likelihood fit and Bayesian model preference is
particularly relevant in light of recent analyses of evolving dark
energy, where the statistical assessment of departures from
$\Lambda$CDM has been shown to depend sensitively on the Bayesian
evidence and the associated prior volume \cite{Ong:2025utx}.
The same distinction is reflected by the information criteria:
Model 2 gives $\Delta{\rm AIC}=0.979$, whereas the stronger complexity
penalty of the BIC yields $\Delta{\rm BIC}=17.330$.
Model 1 is much more strongly disfavored. Its global minimum,
\begin{equation}
\chi^2_{\min}=1583.527,
\end{equation}
is already larger than that of $\Lambda$CDM by
$\Delta\chi^2=62.958$. The corresponding evidence difference,
\begin{equation}
\Delta\ln Z_{\rm KHDE1}
=
-31.929\pm0.213,
\end{equation}
constitutes strong Bayesian evidence against Model 1 for the data combination
and priors considered here. The same conclusion follows independently
from both information criteria, with
$\Delta{\rm AIC}=64.958$ and
$\Delta{\rm BIC}=70.408$.
The markedly different behavior of the two Kaniadakis realizations is
particularly informative. In Model 1, the normalization condition
links the Kaniadakis contribution to $c^2$ and $\Omega_{m0}$, producing
a restricted background evolution. Indeed, after eliminating
$\kappa$, the normalized expansion history can be written in terms of
the combination
\begin{equation}
r=\frac{\Omega_{m0}}{1-c^2},
\end{equation}
which exposes the strong background degeneracy between $c^2$ and
$\Omega_{m0}$. The combined SN and BAO measurements strongly constrain
this restricted expansion history, accounting for most of the
increase in $\chi^2$ found for Model 1.
By contrast, the $\beta\dot H$ contribution of Model 2 breaks this
restricted background structure and provides the additional freedom
required to accommodate the combined low-redshift observations.
Therefore, the derivative term is not merely a phenomenological
extension of Model 1: it has a direct impact on the observational
viability of the Kaniadakis construction. Current observations,
however, do not yet provide sufficient statistical support for this
additional freedom to overcome the Bayesian Occam penalty.
Overall, the Bayesian hierarchy obtained from the combined data is
\begin{equation}
\Lambda{\rm CDM}
>
{\rm HDE}
>
{\rm CPL}
>
{\rm KHDE2}
\gg
{\rm KHDE1},
\end{equation}
where the ordering refers to the Bayesian evidence. The result should
not be interpreted as an observational exclusion of the Kaniadakis
framework as a whole. Rather, it shows that the minimal realization
represented by Model 1 is strongly disfavored, whereas Model 2 remains
capable of providing a competitive fit to the data but is not
currently preferred once its larger parameter space is taken into
account.}
{It is instructive to compare these results with recent applications
of DESI DR2 data to entropic cosmologies. In the Kaniadakis
holographic dark-energy analysis of Ref.~\cite{Luciano:2025ykr},
late-time data including PantheonPlus, cosmic chronometers, and DESI
DR2 were found to provide competitive fits with respect to
$\Lambda$CDM, although the latter remained slightly favored according
to the Akaike information criterion. Our analysis extends this
comparison by including RSD growth measurements and, more
importantly, by computing the Bayesian evidence over the full prior
volume. For the minimal Kaniadakis realization considered here, this
leads to a substantially stronger statistical preference for
$\Lambda$CDM.
A similar qualitative tendency has been reported for other generalized
entropic cosmologies. Barrow- and Tsallis-based scenarios constrained
with DESI DR2 can depart from their standard limits while remaining
slightly disfavored relative to $\Lambda$CDM once information criteria
and Bayesian evidence are taken into account
\cite{Luciano:2025hjn}. Conversely, the recently
proposed Luciano--Saridakis entropic cosmology was found to favor a
departure from the $\Lambda$CDM limit at the $2\sigma$ level when
DESI DR2 is combined with PantheonPlus+SH0ES, cosmic chronometers,
and compressed Planck information
\cite{Leizerovich:2026pfy}.
These comparisons emphasize that DESI DR2 does not generically favor
or disfavor entropic modifications of cosmology. The statistical
outcome depends on the functional form of the entropy-induced
correction, the associated parameter space, and the combination of
cosmological probes employed. In the present Kaniadakis construction,
the distinction between Models 1 and 2 is particularly illustrative:
the minimal realization is strongly disfavored, whereas the
$\beta\dot H$ extension achieves a better maximum-likelihood fit than
$\Lambda$CDM but does not overcome the Bayesian penalty associated
with its additional parameter volume.}

\begin{table*}[t]
\centering
\caption{{Bayesian and information-criterion comparison for the five cosmological
models using the combined CC+PantheonPlus+DESI DR2+RSD data set.
The quantities $\Delta\chi^2$, $\Delta{\rm AIC}$,
$\Delta{\rm BIC}$, and $\Delta\ln Z$ are defined relative to
$\Lambda$CDM. Negative $\Delta\ln Z$ indicates Bayesian preference
for $\Lambda$CDM.}}

\label{tab:bayesian_comparison}
\begin{ruledtabular}
\begin{tabular}{lccccccccc}
Model
& $k$
& $\chi^2_{\min}$
& $\Delta\chi^2$
& AIC
& $\Delta$AIC
& BIC
& $\Delta$BIC
& $\ln Z$
& $\Delta\ln Z$
\\
\hline
$\Lambda$CDM
& 4
& 1520.569
& 0
& 1528.569
& 0
& 1550.369
& 0
& $-776.450\pm0.151$
& 0
\\
HDE
& 5
& 1520.131
& $-0.438$
& 1530.131
& 1.562
& 1557.381
& 7.012
& $-778.831\pm0.160$
& $-2.381\pm0.220$
\\
CPL
& 6
& 1519.380
& $-1.189$
& 1531.380
& 2.811
& 1564.080
& 13.711
& $-780.160\pm0.166$
& $-3.710\pm0.224$
\\
KHDE2
& 7
& 1515.548
& $-5.021$
& 1529.548
& 0.979
& 1567.699
& 17.330
& $-780.807\pm0.174$
& $-4.357\pm0.230$
\\
KHDE1
& 5
& 1583.527
& 62.958
& 1593.527
& 64.958
& 1620.777
& 70.408
& $-808.379\pm0.151$
& $-31.929\pm0.213$
\end{tabular}
\end{ruledtabular}
\end{table*}


\section{Concluding remarks}\label{sec10}
In this work, we have developed a comprehensive analysis of the cosmological and thermodynamic consequences of a holographic formulation of dark energy based on the Kaniadakis entropic deformation. By incorporating this generalization of the Bekenstein-Hawking entropy within a spatially flat FLRW background, we demonstrate that the dark sector acquires a distinctive infrared correction proportional to $H^{-2}$, which substantially modifies the dynamics of the apparent horizon and, by extension, the thermodynamic structure of the Universe.
One of the most relevant results of the study is the derivation of an effective dark energy density of the form \eqref{eq7}
where the infrared term emerges directly from the structure of the warped entropy and not from an ad hoc geometric choice. This result establishes an explicit connection between relativistic non-extensive statistics and late cosmological dynamics, providing a physically transparent way to introduce modifications to the standard holographic scenario.
From a thermodynamic point of view, the formulation based on the Hayward-Kodama formalism and the unified first law allowed the derivation of a geometric equation of state for the apparent horizon. Criticality analysis reveals the existence of a well-defined critical point and a Van der Waals-type structure; however, the system's behavior exhibits a highly unconventional feature: the appearance of an inverted first-order phase transition at temperatures above the critical temperature. This feature is accompanied by a swallowtail-type structure in the Gibbs free energy that does not reach a thermodynamic minimum, indicating the presence of unstable branches and suggesting limits to the equilibrium interpretation for dynamical cosmological horizons.
Extending the model with a dynamical term proportional to $\dot{H}$—inspired by the Granda-Oliveros prescription—does not alter the qualitative nature of these phenomena, but it intensifies the deviations from standard thermodynamic behavior. This result is particularly significant, as it indicates that the critical structure is not an artifact of a specific parameterization, but rather a robust consequence of entropic deformation.
The classical stability study allowed us to establish a lower bound for the dark energy density by requiring a positive effective speed of sound. This condition introduces clear physical constraints on the parameter space and strengthens the internal consistency of the model, showing that infrared modification can sustain dynamically stable configurations within a realistic cosmological regime.

{From an observational standpoint, we confronted both Kaniadakis
realizations with cosmic chronometers, PantheonPlus Type Ia
supernovae, DESI DR2 baryon acoustic oscillations, and redshift-space
distortion measurements. The resulting constraints reveal
characteristic degeneracy directions whose structure depends
sensitively on the model specification. More importantly, the
Bayesian model comparison against $\Lambda$CDM, CPL, and standard
holographic dark energy leads to markedly different conclusions for
the two Kaniadakis realizations. Model 1 is strongly disfavored by the
combined data, with
$\Delta\ln Z=-31.929\pm0.213$ relative to $\Lambda$CDM, consistently
with its substantially larger minimum chi-square. This result can be
traced to the restricted background evolution induced by the exact
relation among $c^2$, $\Omega_{m0}$, and the Kaniadakis contribution.
Model 2 behaves qualitatively differently. The additional
$\beta\dot H$ term relaxes the restricted background structure of
Model 1 and yields the lowest minimum chi-square among all five
cosmologies considered, with $\Delta\chi^2=-5.021$ relative to
$\Lambda$CDM. Nevertheless, the corresponding Bayesian evidence,
$\Delta\ln Z=-4.357\pm0.230$, shows that the improvement in
maximum-likelihood fit is not sufficient to compensate for the larger
parameter space. Thus, current data do not statistically favor the
extended Kaniadakis model over $\Lambda$CDM, although they demonstrate
that the derivative contribution has a direct and significant impact
on the observational viability of the construction.}


{Beyond these degeneracies, a particularly relevant outcome of 
the joint analysis is that both Kaniadakis models naturally 
reproduce a dynamical dark energy equation of state, including 
phantom-like behavior and an effective crossing of the $w=-1$ 
line-without invoking a fundamental phantom field or any phenomenological parameterization such as the CPL ansatz. 
This qualitative trend coincides with the preference for 
phantom crossing reported by the DESI DR2 extended dark energy 
analysis and confirmed by non-parametric reconstructions of 
$w(z)$, suggesting that the Kaniadakis framework provides a 
theoretically motivated alternative to purely phenomenological 
descriptions of late-time cosmic acceleration.}

Taken together, the results position Kaniadakis's entropy-based 
holographic cosmology as a theoretical framework capable of 
linking gravitational thermodynamics, generalized statistical 
mechanics, and dark energy dynamics within a coherent structure. 
Furthermore, the infrared term introduces a late sensitivity 
absent in other entropic generalizations, potentially offering 
new insights into the microphysical mechanisms responsible for 
cosmic acceleration.
{Among the main applications and extensions of this work are: 
(i) the inclusion of high-redshift information from CMB 
anisotropies, which is expected to probe the parameter 
combinations orthogonal to the degenerate directions identified 
here and to tighten the constraints on the deformation 
parameters individually; (ii) stability analysis beyond the 
classical regime, incorporating criteria of causality and 
scale-dependent perturbations; and (iii) the exploration of 
non-equilibrium thermodynamic formulations that allow for a 
deeper interpretation of the nature of the observed inverted 
transitions.}

One of the main findings of this work is that the three phenomena studied in this manuscript have a common origin, namely a single organizing principle, the two-scale structure of the Kaniadakis entropy $S_K=S_{\rm BH}+K^2S_{\rm BH}^3/6+\mathcal{O}(K^4)$. The argument can be formulated systematically as follows. The two terms in $S_K$ lead to the effective dark-energy density $\rho_{de}=3c^2H^2+K^2/H^2$. The structure generates an equation of state for the apparent horizon, Eq.~(24), where the term $K^2v^2/4$ appears with the opposite sign in contrast to a standard Van der Waals system, because the IR correction $K^2/H^2$ grows with the size of the horizon. The sign reversal reverses the phase transition and the swallowtail structure of the Gibbs free energy. This is not due to some particular parameterization but follows from the positivity of the Kaniadakis correction $K^2S_{BH}^3/6>0$.  The dark-energy perturbation $\delta\rho_{de}=\rho_{de}\,A(z)\,\delta H/H$ has the sign of $A(z)=(3c^2H^2-K^2/H^2)/\rho_{de}$, which takes into account the entropic balance of the two Kaniadakis terms since it $\rho_{de}$ is a function of $H$ both positive and negative powers. The perturbative analogue of the thermodynamic transition is the zero crossing $A(z_*)=0$, where the effective gravitational coupling crosses its GR value. Both conditions reduce to $3c^2H^2=K^2/H^2$, the same equation that divides the UV- and IR-dominated regimes of $S_K$. This same competition determines the expansion history $H(z)$ via the modified Friedmann equation and thus fixes the epoch at which the IR correction becomes cosmologically relevant. The late-time deviation $H(z)$ from $\Lambda$CDM, the zero crossing of it, $A(z)$ and the inverted thermodynamic phase transition all occur in the interval $z\lesssim 1$ exactly because this is where $K^2/H^2\sim 3c^2H^2$ along the background trajectory compatible with the observational data of Tables~I-III.
The three phenomena are thus not independent signatures but three projections of the same underlying entropic structure on the thermodynamic, perturbative and background sectors of the theory.
We note that breaking the remaining parameter degeneracies, namely the $c^2$--$\Omega_{m0}$ degeneracy in Model~1 and the $c^2$--$K^2/H_0^4$ anticorrelation in Model 2, will require complementary high-redshift information such as CMB anisotropies or independent priors on $\sigma_{8,0}$, and is a natural direction for future work.
{In summary, the Kaniadakis entropic deformation provides a unified
framework in which the same effective ultraviolet-infrared competition
is manifested in the background evolution, perturbative dynamics, and
thermodynamics of the apparent horizon. The observational analysis,
however, shows that the viability of this construction depends
strongly on its cosmological realization. The minimal model is
strongly disfavored by current late-time data, whereas the extended
model remains competitive at the level of the maximum likelihood but
is not preferred over $\Lambda$CDM by Bayesian evidence. This
distinction identifies the derivative sector as an essential
ingredient for phenomenological viability rather than a merely
optional extension. Future constraints incorporating high-redshift
information, particularly CMB anisotropies, will be necessary to
determine whether the additional dynamical freedom can be supported
observationally and to further test the connection between entropic
deformations, horizon thermodynamics, and cosmic acceleration.}

\section*{Acknowledgements}

We thank the anonymous referee for valuable comments and suggestions. S. Lepe acknowledges the financial support of Fondecyt Grant No.1250969 and
J. Saavedra acknowledges the financial support of Fondecyt Grant 1220065.

\section*{Conflict of Interest}

The authors declare no conflict of interest.

\section*{Data Availability Statement}

Data sharing is not applicable to this article as no new data were created or analyzed in this study.

\section*{Keywords}

Cosmology; Kaniadakis entropy; Dark energy; Observational data.

\appendix

\section{Cosmological dynamics with generalized holographic dark energy}\label{app:A}

We consider a generalized holographic dark energy density of the form
\begin{equation}
\rho_{\mathrm{de}} = 3c^{2}H^{2} + \frac{K^{2}}{H^{2}},
\end{equation}
where $c^{2}$ and $K$ are constant parameters.

The Friedmann equations for a spatially flat universe ($k=0$) read
\begin{align}
3H^{2} &= \rho + \rho_{\mathrm{de}}, \\
\dot{\rho} + 3H(\rho + p) &= 0, \qquad
\dot{\rho}_{\mathrm{de}} + 3H(\rho_{\mathrm{de}} + p_{\mathrm{de}}) = 0, \\
-2\dot{H} &= \rho + p + \rho_{\mathrm{de}} + p_{\mathrm{de}}.
\end{align}

Substituting the expression for $\rho_{\mathrm{de}}$ into the Friedmann constraint, we obtain an algebraic equation for $H(z)$,
\begin{equation}\label{eqA5}
H^{2}(z) = \frac{1}{6(1-c^{2})}
\left[
1 + \sqrt{1 + \frac{12(1-c^{2})K^{2}}{\rho^{2}(z)}}
\right]\rho(z).
\end{equation}

An interesting feature of this model is the existence of a self-accelerating solution even in the absence of standard matter, i.e., for $\rho \to 0$, provided that $c^{2}<1$. In this limit, the Hubble parameter approaches a constant value,
\begin{equation}
H = \left(\frac{3}{1-c^{2}}\right)^{1/4}\sqrt{\frac{K}{3}}.
\end{equation}

Assuming a barotropic equation of state $p=\omega \rho$ with constant $\omega$, the matter density evolves as
\begin{equation}
\rho(z) = \rho_{0}(1+z)^{3(1+\omega)}.
\end{equation}
For $\omega > -1$, one has $\rho(z\to -1)\to 0$, and therefore
\begin{equation}\label{A8}
H(z\to -1) \to \left(\frac{3}{1-c^{2}}\right)^{1/4}\sqrt{\frac{K}{3}}.
\end{equation}

This asymptotic behavior resembles that of the $\Lambda$CDM model,
\begin{equation}
3H^{2}(z) = \rho_{0}(1+z)^{3} + \Lambda,
\end{equation}
for which $H(z\to -1) \to \sqrt{\Lambda/3}$. This suggests the identification
\begin{equation}
K = \sqrt{\frac{1-c^{2}}{3}}\,\Lambda,
\end{equation}
which provides an effective interpretation of the parameter $K$ in terms of a cosmological constant.

\paragraph{Deceleration parameter.}

Using the Friedmann equations, the deceleration parameter is given by
\begin{equation}
q = -1 + \frac{3}{2}
\left(
1 + \frac{p + p_{\mathrm{de}}}{\rho + \rho_{\mathrm{de}}}
\right)
= \frac{1}{2}\left[1 + 3\left(\omega \Omega + \omega_{\mathrm{de}}\Omega_{\mathrm{de}}\right)\right],
\end{equation}
where we have introduced the density parameters
\begin{equation}
\Omega = \frac{\rho}{3H^{2}}, \qquad
\Omega_{\mathrm{de}} = \frac{\rho_{\mathrm{de}}}{3H^{2}} = c^{2} + \frac{K^{2}}{3H^{4}}.
\end{equation}

From the above expression, one obtains
\begin{equation}
\omega \Omega + \omega_{\mathrm{de}}\Omega_{\mathrm{de}} = \frac{1}{3}(2q - 1).
\end{equation}

In the case of cold dark matter ($\omega = 0$), the deceleration parameter constrains the dark energy equation of state. In particular:

\begin{itemize}
\item For quintessence ($-1 < \omega_{\mathrm{de}} < -1/3$),
\begin{equation}
-\frac{1}{2}\left(3\Omega_{\mathrm{de}} + 1\right)
< q <
\frac{1}{2}\left(1 - \Omega_{\mathrm{de}}\right).
\end{equation}

\item For phantom dark energy ($\omega_{\mathrm{de}} < -1$),
\begin{equation}
q < -\frac{1}{2}\left(3\Omega_{\mathrm{de}} - 1\right).
\end{equation}
\end{itemize}

These relations may provide observational insight into the present value of $\Omega_{\mathrm{de}}$, which in this model reads
\begin{equation}
\Omega_{\mathrm{de}}(0) = c^{2} + \frac{K^{2}}{3H_{0}^{4}}.
\end{equation}
This raises the question of whether observations can independently constrain both parameters $c^{2}$ and $K$.

\paragraph{Including $\dot{H}$ corrections.}

A natural extension of the model consists in adding a term proportional to $\dot{H}$ in the dark energy density,
\begin{equation}
\rho_{\mathrm{de}} = 3c^{2}H^{2} + \frac{K^{2}}{H^{2}} + 3\beta \dot{H},
\end{equation}
where $\beta$ is a constant parameter.

Substituting this expression into the Friedmann constraint leads to the differential equation
\begin{equation}
\frac{dH^{2}}{dz}
+ \frac{2}{\beta}(1-c^{2})\frac{H^{2}}{1+z}
=
\frac{2}{3\beta}
\left(\rho + \frac{K^{2}}{H^{2}}\right)
\frac{1}{1+z},
\end{equation}
where we have used the relation
\begin{equation}
\dot{H} = -\frac{1}{2}(1+z)\frac{dH^{2}}{dz}.
\end{equation}

\section{Effective Dark-Energy Equation of State and Regimes}
\label{app:B}

The effective dark-energy equation-of-state parameter can be written in the form
\begin{align}
\omega_{de}
=
-1
+
(1+\omega)\,
\frac{A\,r}{1+r-A},
\tag{B1}\label{eq:B1}
\end{align}
where the auxiliary quantity $A$ is defined as
\begin{align}
A
=
\frac{3c^{2}H^{2}-K^{2}/H^{2}}
     {3c^{2}H^{2}+K^{2}/H^{2}}.
\tag{B2}\label{eq:B2}
\end{align}
Here $r$ denotes the ratio of energy densities (as defined in the main text), and $\omega$ is the equation-of-state parameter of the matter sector.

\subsection*{Sign of $A$ and corresponding dynamical regimes}

From Eq.~\eqref{eq:B2} it follows that
\begin{align}
0 < A < 1
\quad &\to \quad
H^{2} > \frac{1}{\sqrt{3}\,c}\,K,
\tag{B3}\label{eq:B3}
\\
A < 0
\quad &\to \quad
H^{2} < \frac{1}{\sqrt{3}\,c}\,K.
\tag{B4}\label{eq:B4}
\end{align}
If the condition \eqref{eq:B4} holds ($A<0$), then Eq.~\eqref{eq:B1} becomes
\begin{align}
\omega_{de}
=
-1
-
(1+\omega)\,
\frac{|A|\,r}{1+r+|A|}
\qquad
\text{(phantom regime)}.
\tag{B5}\label{eq:B5}
\end{align}
If instead \eqref{eq:B3} holds ($0<A<1$), then $\omega_{de}$ may lie in a quintessence-like range,
\begin{align}
-1 < \omega_{de} < -\frac13
\quad \to \quad
0 < (1+\omega)\,\frac{A r}{1+r-A} < \frac{2}{3},
\tag{B6}\label{eq:B6}
\end{align}
or it may satisfy
\begin{align}
\omega_{de} > 0
\quad \to \quad
(1+\omega)\,\frac{A r}{1+r-A} > 1.
\tag{B7}\label{eq:B7}
\end{align}
From the standpoint of interpreting $\rho_{de}$ as an effective dark-energy component driving cosmic acceleration, the case \eqref{eq:B7} (i.e.\ $\omega_{de}>0$) is typically regarded as physically implausible for late-time acceleration. Therefore, the phenomenologically relevant regimes are usually those described by either the phantom condition \eqref{eq:B5} or the quintessence-like condition \eqref{eq:B6}.

\section{The Case $\beta = 0$ and Explicit Expressions for $q(z)$ and $T_A(z)$}
\label{app:C}

Setting $\beta=0$, the Hubble function simplifies to
\begin{widetext}
\begin{align}
H^{2}(z)
=
\frac{H_{0}^{2}}{2(1-c^{2})}
\left[
\Omega(z)
+
\sqrt{
\Omega^{2}(z)
+
\frac{4(1-c^{2})K^{2}}{3H_{0}^{2}}
}
\right].
\tag{C1}\label{eq:C1}
\end{align}
For cold dark matter ($\omega=0$), one has $\Omega(z)=\Omega_{0}(1+z)^{3}$, and Eq.~\eqref{eq:C1} becomes
\begin{align}
H^{2}(z)
=
\frac{H_{0}^{2}}{2(1-c^{2})}
\left[
\Omega_{0}(1+z)^{3}
+
\sqrt{
\Omega_{0}^{2}(1+z)^{6}
+
\frac{4(1-c^{2})K^{2}}{3H_{0}^{2}}
}
\right].
\tag{C2}\label{eq:C2}
\end{align}
\end{widetext}
\subsection*{Apparent-horizon temperature}

The apparent-horizon temperature can be expressed as
\begin{align}
T_{A}(z)
=
\frac{H(z)}{4\pi}\,[1-q(z)],
\qquad
q=-1 \ \to\
T_{A}(z)=\frac{H(z)}{2\pi}.
\tag{C3}\label{eq:C3}
\end{align}
Thus, when the background approaches a de Sitter stage ($q\to -1$), the temperature reduces to the Kodama--Hayward form shown in Eq.~\eqref{eq:C3}.

\subsection*{From time derivatives to redshift derivatives}

Using $dz/dt = -(1+z)H$, one may rewrite $\dot H$ in terms of derivatives with respect to $z$ as
\begin{align}
\dot{H}
=
-\frac12(1+z)\,\frac{dH^{2}}{dz}.
\tag{C4}\label{eq:C4}
\end{align}
Since the deceleration parameter is $q=-1-\dot H/H^{2}$, this becomes
\begin{align}
q(z)
=
-1
+
\frac12(1+z)\,
\frac{1}{H^{2}}\frac{dH^{2}}{dz}.
\tag{C5}\label{eq:C5}
\end{align}

Differentiating Eq.~\eqref{eq:C2} with respect to $z$ yields
\begin{widetext}
\begin{align}
\frac{dH^{2}}{dz}
=
\frac{3H_{0}^{2}\Omega_{0}}{2(1-c^{2})}
\left(
1+
\frac{\Omega_{0}(1+z)^{3}}
{\sqrt{
\Omega_{0}^{2}(1+z)^{6}
+
4(1-c^{2})K^{2}/(3H_{0}^{2})
}}
\right)
(1+z)^{2}.
\tag{C6}\label{eq:C6}
\end{align}
\end{widetext}
Substituting Eq.~\eqref{eq:C6} into Eq.~\eqref{eq:C5}, one obtains the explicit deceleration parameter
\begin{align}
q(z)
=
-1
+
\frac{3}{2}
\left(
\frac{\Omega_{0}(1+z)^{3}}
{\sqrt{
\Omega_{0}^{2}(1+z)^{6}
+
4(1-c^{2})K^{2}/(3H_{0}^{2})
}}
\right).
\tag{C7}\label{eq:C7}
\end{align}
From Eq.~\eqref{eq:C7} it follows immediately that
\begin{align}
q(z\gg 0) \sim \frac12,
\qquad
q(z\to -1)\to -1,
\tag{C8}\label{eq:C8}
\end{align}
corresponding to an early-time matter-dominated regime ($q=1/2$) and a late-time cosmological-constant-like regime ($q\to -1$).

\subsection*{Early- and late-time limits of $H(z)$ and $T_A(z)$}

From Eq.~\eqref{eq:C2} one finds
\begin{widetext}
\begin{align}
H(z\gg 0)
&\sim
H_{0}\sqrt{\frac{\Omega_{0}}{1-c^{2}}}\,(1+z)^{3},
\qquad
H(z\to -1)
\longrightarrow
\sqrt{
\frac{H_{0}K}{\sqrt{3(1-c^{2})}}
},
\tag{C9}\label{eq:C9}
\end{align}
\end{widetext}
and the late-time limit in Eq.~\eqref{eq:C9} coincides with the limit found in Eq.~\eqref{A8}.

Using Eq.~\eqref{eq:C3}, the temperature behaves as
\begin{widetext}
\begin{align}
T_{A}(z\gg 0)
&\sim
\frac{H_{0}}{8\pi}\sqrt{\frac{\Omega_{0}}{1-c^{2}}}\,(1+z)^{3},
\qquad
T_{A}(z\to -1)
\longrightarrow
\frac{1}{2\pi}
\sqrt{
\frac{H_{0}K}{\sqrt{3(1-c^{2})}}
}.
\tag{C10}\label{eq:C10}
\end{align}
\end{widetext}

\subsection*{Present-day constraint and parameter estimates}

Evaluating Eq.~\eqref{eq:C7} at $z=0$ gives
\begin{align}
q(0)
=
-1
+
\frac{3}{2}
\left(
\frac{\Omega_{0}}
{\sqrt{
\Omega_{0}^{2}
+
4(1-c^{2})K^{2}/(3H_{0}^{2})
}}
\right).
\tag{C11}\label{eq:C11}
\end{align}

As a simple estimate, if one adopts $q(0)\simeq -0.55$ and $\Omega_{0}\simeq 0.3$, then $q(0)\simeq -1/2$ suggests a solution
\begin{align}
\frac{K}{H_{0}}
\sim
\sqrt{\frac{6}{1-c^{2}}}\,\Omega_{0},
\tag{C12}\label{eq:C12}
\end{align}
which still contains two unknowns, $c^{2}$ and $K$.

From Eq.~\eqref{eqA5} with $\beta=0$, one obtains at $z=0$ the exact relation
\begin{widetext}
\begin{align}
1
=
\frac{\Omega_{0}}{2(1-c^{2})}
\left[
1+
\sqrt{
1+
\frac{4(1-c^{2})K^{2}}{3H_{0}^{2}\Omega_{0}^{2}}
}
\right]
\quad \Longrightarrow \quad
\frac{K}{H_{0}}
=
\sqrt{3\left[1-c^{2}-\Omega_{0}\right]}.
\tag{C13}\label{eq:C13}
\end{align}
\end{widetext}
Combining Eqs.~\eqref{eq:C12} and \eqref{eq:C13} yields
\begin{align}
\sqrt{\frac{6}{1-c^{2}}}\,\Omega_{0}
\sim
\sqrt{3\left[1-c^{2}-\Omega_{0}\right]},
\tag{C14}\label{eq:C14}
\end{align}
which implies
\begin{align}
\left(\frac{2\Omega_{0}}{1-c^{2}}+1\right)\Omega_{0}
\sim
1-c^{2}
\quad \Longrightarrow \quad
c^{2}
\sim
1-2\Omega_{0}.
\tag{C15}\label{eq:C15}
\end{align}
Substituting Eq.~\eqref{eq:C15} into Eq.~\eqref{eq:C13} then gives
\begin{align}
K
\sim
\sqrt{3\Omega_{0}}\;H_{0}.
\tag{C16}\label{eq:C16}
\end{align}
Therefore, one arrives at the rough estimates
\begin{align}
c^{2}\sim 1-2\Omega_{0},
\qquad
K\sim \sqrt{3\Omega_{0}}\;H_{0},
\tag{C17}\label{eq:C17}
\end{align}
showing explicitly how observations of $\Omega_{0}$ and $H_{0}$ could be used to infer $(c^{2},K)$ at the level of order-of-magnitude estimates.

Finally, if one compares the late-time limit in Eq.~\eqref{eq:C9} with the $\Lambda$CDM expectation $H(z\to -1)\to \sqrt{\Lambda/3}$, one obtains
\begin{align}
H(z\to -1)
\longrightarrow
\sqrt{
\frac{H_{0}K}{\sqrt{3(1-c^{2})}}
}
\sim
\sqrt{\frac{\Lambda}{3}},
\tag{C18}\label{eq:C18}
\end{align}
which suggests the scaling
\begin{align}
H_{0}K
\sim
\sqrt{\frac{1}{3}(1-c^{2})}\,\Lambda,
\tag{C19}\label{eq:C19}
\end{align}
and using Eq.~\eqref{eq:C15} this may be expressed as
\begin{align}
K
\sim
\sqrt{\frac{2\Omega_{0}}{3}}\,
\frac{\Lambda}{H_{0}}.
\tag{C20}\label{eq:C20}
\end{align}

\bibliography{biblio.bib}

@article{DESI:2025wyn,
    author = "Gu, Gan and others",
    collaboration = "DESI",
    title = "{Dynamical dark energy in light of the DESI DR2 baryonic acoustic oscillations measurements}",
    eprint = "2504.06118",
    archivePrefix = "arXiv",
    primaryClass = "astro-ph.CO",
    reportNumber = "FERMILAB-PUB-25-0235-PPD",
    doi = "10.1038/s41550-025-02669-6",
    journal = "Nature Astron.",
    volume = "9",
    number = "12",
    pages = "1879--1889",
    year = "2025",
    note = "[Erratum: Nature Astron. 9, 1898 (2025)]"
}

@article{DESI:2025fii,
    author = "Lodha, K. and others",
    collaboration = "DESI",
    title = "{Extended dark energy analysis using DESI DR2 BAO measurements}",
    eprint = "2503.14743",
    archivePrefix = "arXiv",
    primaryClass = "astro-ph.CO",
    reportNumber = "FERMILAB-PUB-25-0164-PPD",
    doi = "10.1103/w4c6-1r5j",
    journal = "Phys. Rev. D",
    volume = "112",
    number = "8",
    pages = "083511",
    year = "2025"
}

@article{Linder:2002et,
    author = "Linder, Eric V.",
    title = "{Exploring the expansion history of the universe}",
    eprint = "astro-ph/0208512",
    archivePrefix = "arXiv",
    doi = "10.1103/PhysRevLett.90.091301",
    journal = "Phys. Rev. Lett.",
    volume = "90",
    pages = "091301",
    year = "2003"
}

@article{Chevallier:2000qy,
    author = "Chevallier, Michel and Polarski, David",
    title = "{Accelerating universes with scaling dark matter}",
    eprint = "gr-qc/0009008",
    archivePrefix = "arXiv",
    doi = "10.1142/S0218271801000822",
    journal = "Int. J. Mod. Phys. D",
    volume = "10",
    pages = "213--224",
    year = "2001"
}

@article{Kazantzidis:2018rnb,
      author         = "Kazantzidis, Lavrentios and Perivolaropoulos, Leandros",
      title          = "{Evolution of the $f\sigma_8$ tension with the
                        Planck15/$\Lambda$CDM determination and implications for
                        modified gravity theories}",
      journal        = "Phys. Rev.",
      volume         = "D97",
      year           = "2018",
      number         = "10",
      pages          = "103503",
      doi            = "10.1103/PhysRevD.97.103503",
      eprint         = "1803.01337",
      archivePrefix  = "arXiv",
      primaryClass   = "astro-ph.CO",
      SLACcitation   = "%%CITATION = ARXIV:1803.01337;%%"
}

@article{Planck:2018vyg,
    author = "Aghanim, N. and others",
    collaboration = "Planck",
    title = "{Planck 2018 results. VI. Cosmological parameters}",
    eprint = "1807.06209",
    archivePrefix = "arXiv",
    primaryClass = "astro-ph.CO",
    doi = "10.1051/0004-6361/201833910",
    journal = "Astron. Astrophys.",
    volume = "641",
    pages = "A6",
    year = "2020",
    note = "[Erratum: Astron.Astrophys. 652, C4 (2021)]"
}

@article{Brout:2022vxf,
    author = "Brout, Dillon and others",
    title = "{The Pantheon+ Analysis: Cosmological Constraints}",
    eprint = "2202.04077",
    archivePrefix = "arXiv",
    primaryClass = "astro-ph.CO",
    doi = "10.3847/1538-4357/ac8e04",
    journal = "Astrophys. J.",
    volume = "938",
    number = "2",
    pages = "110",
    year = "2022"
}

@article{Scolnic:2021amr,
    author = "Scolnic, Dan and others",
    title = "{The Pantheon+ Analysis: The Full Data Set and Light-curve Release}",
    eprint = "2112.03863",
    archivePrefix = "arXiv",
    primaryClass = "astro-ph.CO",
    doi = "10.3847/1538-4357/ac8b7a",
    journal = "Astrophys. J.",
    volume = "938",
    number = "2",
    pages = "113",
    year = "2022"
}

@article{Brownsberger:2021uue,
    author = "Brownsberger, Sasha R. and Brout, Dillon and Scolnic, Daniel and Stubbs, Christopher W. and Riess, Adam G.",
    title = "{Dependence of Cosmological Constraints on Gray Photometric Zero-point Uncertainties of Supernova Surveys}",
    eprint = "2110.03486",
    archivePrefix = "arXiv",
    primaryClass = "astro-ph.CO",
    doi = "10.3847/1538-4357/acad80",
    journal = "Astrophys. J.",
    volume = "944",
    number = "2",
    pages = "188",
    year = "2023"
}

@article{DESI:2025zpo,
    author = "Abdul Karim, M. and others",
    collaboration = "DESI",
    title = "{DESI DR2 results. I. Baryon acoustic oscillations from the Lyman alpha forest}",
    eprint = "2503.14739",
    archivePrefix = "arXiv",
    primaryClass = "astro-ph.CO",
    reportNumber = "FERMILAB-PUB-25-0167-PPD",
    doi = "10.1103/2wwn-xjm5",
    journal = "Phys. Rev. D",
    volume = "112",
    number = "8",
    pages = "083514",
    year = "2025"
}

@article{DESI:2025zgx,
    author = "Abdul Karim, M. and others",
    collaboration = "DESI",
    title = "{DESI DR2 results. II. Measurements of baryon acoustic oscillations and cosmological constraints}",
    eprint = "2503.14738",
    archivePrefix = "arXiv",
    primaryClass = "astro-ph.CO",
    reportNumber = "FERMILAB-PUB-25-0169-PPD",
    doi = "10.1103/tr6y-kpc6",
    journal = "Phys. Rev. D",
    volume = "112",
    number = "8",
    pages = "083515",
    year = "2025"
}

@article{farooq2013hubble,
  title={Hubble parameter measurement constraints on the cosmological deceleration--acceleration transition redshift},
  author={Farooq, Omer and Ratra, Bharat},
  journal={The Astrophysical Journal Letters},
  volume={766},
  number={1},
  pages={L7},
  year={2013},
  publisher={IOP Publishing}
}

@article{cao2018cosmological,
  title={Cosmological model-independent test of$\backslash$varLambda $\Lambda$ CDM with two-point diagnostic by the observational Hubble parameter data},
  author={Cao, Shu-Lei and Duan, Xiao-Wei and Meng, Xiao-Lei and Zhang, Tong-Jie},
  journal={The European Physical Journal C},
  volume={78},
  pages={1--16},
  year={2018},
  publisher={Springer}
}

@article{Moresco:2020fbm,
    author = "Moresco, Michele and Jimenez, Raul and Verde, Licia and Cimatti, Andrea and Pozzetti, Lucia",
    title = "{Setting the Stage for Cosmic Chronometers. II. Impact of Stellar Population Synthesis Models Systematics and Full Covariance Matrix}",
    eprint = "2003.07362",
    archivePrefix = "arXiv",
    primaryClass = "astro-ph.GA",
    doi = "10.3847/1538-4357/ab9eb0",
    journal = "Astrophys. J.",
    volume = "898",
    number = "1",
    pages = "82",
    year = "2020"
}

@article{Bekenstein1973,
  author = {J. D. Bekenstein},
  title = {Black holes and entropy},
  journal = {Phys. Rev. D},
  volume = {7},
  pages = {2333--2346},
  year = {1973},
  doi = {10.1103/PhysRevD.7.2333}
}

@article{Hawking1975,
  author = {S. W. Hawking},
  title = {Particle creation by black holes},
  journal = {Commun. Math. Phys.},
  volume = {43},
  pages = {199--220},
  year = {1975},
  doi = {10.1007/BF02345020}
}

@article{Kaniadakis2001,
  author = {G. Kaniadakis},
  title = {Non-linear kinetics underlying generalized statistics},
  journal = {Physica A},
  volume = {296},
  pages = {405--425},
  year = {2001},
  doi = {10.1016/S0378-4371(01)00184-4}
}

@article{Kaniadakis2002,
  author = {G. Kaniadakis},
  title = {Statistical mechanics in the context of special relativity},
  journal = {Phys. Rev. E},
  volume = {66},
  pages = {056125},
  year = {2002},
  doi = {10.1103/PhysRevE.66.056125}
}

@article{Li:2004rb,
    author = "Li, Miao",
    title = "{A Model of holographic dark energy}",
    eprint = "hep-th/0403127",
    archivePrefix = "arXiv",
    doi = "10.1016/j.physletb.2004.10.014",
    journal = "Phys. Lett. B",
    volume = "603",
    pages = "1",
    year = "2004"
}

@article{Wang:2016och,
    author = "Wang, Shuang and Wang, Yi and Li, Miao",
    title = "{Holographic Dark Energy}",
    eprint = "1612.00345",
    archivePrefix = "arXiv",
    primaryClass = "astro-ph.CO",
    doi = "10.1016/j.physrep.2017.06.003",
    journal = "Phys. Rept.",
    volume = "696",
    pages = "1--57",
    year = "2017"
}

@article{Moradpour:2020dfm,
    author = "Moradpour, H. and Ziaie, A. H. and Kord Zangeneh, M.",
    title = "{Generalized entropies and corresponding holographic dark energy models}",
    eprint = "2005.06271",
    archivePrefix = "arXiv",
    primaryClass = "gr-qc",
    reportNumber = "80,732",
    doi = "10.1140/epjc/s10052-020-8307-x",
    journal = "Eur. Phys. J. C",
    volume = "80",
    number = "8",
    pages = "732",
    year = "2020"
}

@article{Faraoni:2011hf,
    author = "Faraoni, Valerio",
    title = "{Cosmological apparent and trapping horizons}",
    eprint = "1106.4427",
    archivePrefix = "arXiv",
    primaryClass = "gr-qc",
    doi = "10.1103/PhysRevD.84.024003",
    journal = "Phys. Rev. D",
    volume = "84",
    pages = "024003",
    year = "2011"
}

@article{Hayward:1997jp,
    author = "Hayward, Sean A.",
    title = "{Unified first law of black hole dynamics and relativistic thermodynamics}",
    eprint = "gr-qc/9710089",
    archivePrefix = "arXiv",
    doi = "10.1088/0264-9381/15/10/017",
    journal = "Class. Quant. Grav.",
    volume = "15",
    pages = "3147--3162",
    year = "1998"
}

@article{Kodama:1979vn,
    author = "Kodama, Hideo",
    title = "{Conserved Energy Flux for the Spherically Symmetric System and the Back Reaction Problem in the Black Hole Evaporation}",
    reportNumber = "KUNS-506",
    doi = "10.1143/PTP.63.1217",
    journal = "Prog. Theor. Phys.",
    volume = "63",
    pages = "1217",
    year = "1980"
}

@article{Cai:2006rs,
    author = "Cai, Rong-Gen and Cao, Li-Ming",
    title = "{Unified first law and thermodynamics of apparent horizon in FRW universe}",
    eprint = "gr-qc/0611071",
    archivePrefix = "arXiv",
    doi = "10.1103/PhysRevD.75.064008",
    journal = "Phys. Rev. D",
    volume = "75",
    pages = "064008",
    year = "2007"
}

@article{Cruz:2023xjp,
    author = "Cruz, Miguel and Lepe, Samuel and Saavedra, Joel",
    title = "{A new approach to P{\ensuremath{-}}V phase transitions: Einstein gravity and holographic type dark energy}",
    eprint = "2312.14257",
    archivePrefix = "arXiv",
    primaryClass = "gr-qc",
    doi = "10.1016/j.dark.2024.101580",
    journal = "Phys. Dark Univ.",
    volume = "46",
    pages = "101580",
    year = "2024"
}

@article{Kubiznak:2012wp,
    author = "Kubiznak, David and Mann, Robert B.",
    title = "{P-V criticality of charged AdS black holes}",
    eprint = "1205.0559",
    archivePrefix = "arXiv",
    primaryClass = "hep-th",
    doi = "10.1007/JHEP07(2012)033",
    journal = "JHEP",
    volume = "07",
    pages = "033",
    year = "2012"
}

@article{Granda:2008dk,
    author = "Granda, L. N. and Oliveros, A.",
    title = "{Infrared cut-off proposal for the Holographic density}",
    eprint = "0810.3149",
    archivePrefix = "arXiv",
    primaryClass = "gr-qc",
    doi = "10.1016/j.physletb.2008.10.017",
    journal = "Phys. Lett. B",
    volume = "669",
    pages = "275--277",
    year = "2008"
}

@article{Planck2018,
  author = {Planck Collaboration},
  title = {Planck 2018 results. VI. Cosmological parameters},
  journal = {Astronomy \& Astrophysics},
  volume = {641},
  year = {2020},
  pages = {A6},
  eprint = {1807.06209},
  archivePrefix = {arXiv}
}

@article{Weinberg1989,
  author = {Steven Weinberg},
  title = {The Cosmological Constant Problem},
  journal = {Reviews of Modern Physics},
  volume = {61},
  pages = {1--23},
  year = {1989}
}

@article{Zlatev1999,
  author = {I. Zlatev and L. Wang and P. J. Steinhardt},
  title = {Quintessence, Cosmic Coincidence, and the Cosmological Constant},
  journal = {Physical Review Letters},
  volume = {82},
  pages = {896},
  year = {1999},
  eprint = {astro-ph/9807002}
}

@article{Hooft1993,
  author = {G. 't Hooft},
  title = {Dimensional reduction in quantum gravity},
  journal = {Salamfestschrift},
  year = {1993},
  eprint = {gr-qc/9310026}
}

@article{Susskind1995,
  author = {L. Susskind},
  title = {The World as a hologram},
  journal = {Journal of Mathematical Physics},
  volume = {36},
  year = {1995},
  pages = {6377},
  eprint = {hep-th/9409089}
}

@article{Li2004,
  author = {M. Li},
  title = {A Model of holographic dark energy},
  journal = {Physics Letters B},
  volume = {603},
  pages = {1},
  year = {2004},
  eprint = {hep-th/0403127}
}

@article{Tsallis1988,
  author = {C. Tsallis},
  title = {Possible generalization of Boltzmann–Gibbs statistics},
  journal = {Journal of Statistical Physics},
  volume = {52},
  pages = {479},
  year = {1988}
}

@article{Renyi1961,
  author = {A. Rényi},
  title = {On measures of entropy and information},
  journal = {Proceedings of the Fourth Berkeley Symposium},
  year = {1961}
}

@article{Barrow2020,
  author = {J. D. Barrow},
  title = {The Area of a Rough Black Hole},
  journal = {Physics Letters B},
  volume = {808},
  pages = {135643},
  year = {2020}
}

@article{Drepanou2022,
  author = {N. Drepanou and A. Lymperis and E. N. Saridakis and K. Yesmakhanova},
  title = {Kaniadakis holographic dark energy and cosmology},
  journal = {European Physical Journal C},
  volume = {82},
  pages = {449},
  year = {2022},
  eprint = {2109.09181}
}

@article{Moradpour2020,
  author = {H. Moradpour and A. Ziaie and M. Kord Zangeneh},
  title = {Generalized entropies and corresponding holographic dark energy models},
  journal = {European Physical Journal C},
  volume = {80},
  pages = {732},
  year = {2020},
  eprint = {2005.06271}
}

@article{Kodama1980,
  author = {H. Kodama},
  title = {Conserved Energy Flux for the Spherically Symmetric System},
  journal = {Progress of Theoretical Physics},
  volume = {63},
  pages = {1217},
  year = {1980}
}

@article{Hayward1998,
  author = {S. A. Hayward},
  title = {Unified first law of black hole dynamics and relativistic thermodynamics},
  journal = {Classical and Quantum Gravity},
  volume = {15},
  pages = {3147},
  year = {1998}
}

@article{MisnerSharp1964,
  author = {C. W. Misner and D. H. Sharp},
  title = {Relativistic equations for adiabatic, spherically symmetric gravitational collapse},
  journal = {Physical Review},
  volume = {136},
  pages = {B571},
  year = {1964}
}

@article{GrandaOliveros2008,
  author = {L. Granda and A. Oliveros},
  title = {Infrared cut-off proposal for the holographic density},
  journal = {Physics Letters B},
  volume = {669},
  pages = {275},
  year = {2008},
  eprint = {0810.3149}
}

@article{Riess2022,
  author = {Riess, Adam G. and others},
  title = {A Comprehensive Measurement of the Local Value of the Hubble Constant with 1 km/s/Mpc Uncertainty from the Hubble Space Telescope and the SH0ES Team},
  journal = {The Astrophysical Journal Letters},
  volume = {934},
  number = {1},
  pages = {L7},
  year = {2022},
  doi = {10.3847/2041-8213/ac5c5b},
  eprint = {2112.04510},
  archivePrefix = {arXiv}
}

@article{Alam2021,
  author = {Alam, Shadab and others},
  title = {The {SDSS-IV} extended {Baryon Oscillation Spectroscopic Survey (eBOSS):} 
           final measurement of {H(z)} from the baryon acoustic oscillation 
           measurements between redshifts 0.6 and 3.5},
  journal = {Monthly Notices of the Royal Astronomical Society},
  volume = {506},
  number = {1},
  pages = {951--975},
  year = {2021},
  doi = {10.1093/mnras/stab1811},
  eprint = {2007.08991},
  archivePrefix = {arXiv}
}

@article{Jacobson:1995ab,
    author = "Jacobson, Ted",
    title = "{Thermodynamics of space-time: The Einstein equation of state}",
    eprint = "gr-qc/9504004",
    archivePrefix = "arXiv",
    reportNumber = "UMDGR-95-114",
    doi = "10.1103/PhysRevLett.75.1260",
    journal = "Phys. Rev. Lett.",
    volume = "75",
    pages = "1260--1263",
    year = "1995"
}

@article{Moresco2016,
  author = {Moresco, M. et al.},
  title = {A 6\% measurement of the Hubble parameter at z~0.45: direct evidence of the epoch of cosmic re-acceleration},
  journal = {JCAP},
  volume = {05},
  pages = {014},
  year = {2016}
}

@article{Ratsimbazafy2017,
  author = {Ratsimbazafy, A. L. et al.},
  title = {Age-dating Luminous Red Galaxies observed with the Southern African Large Telescope},
  journal = {MNRAS},
  volume = {467},
  pages = {3239},
  year = {2017}
}

@article{Brout2022,
  author = {Brout, D. et al.},
  title = {The Pantheon+ Analysis: Cosmological Constraints},
  journal = {ApJ},
  volume = {938},
  pages = {110},
  year = {2022}
}

@article{DESI2024,
    author = "Adame, A. G. et al.",
    collaboration = "DESI",
    title = "{DESI 2024 VI: cosmological constraints from the measurements of baryon acoustic oscillations}",
    eprint = "2404.03002",
    archivePrefix = "arXiv",
    primaryClass = "astro-ph.CO",
    reportNumber = "FERMILAB-PUB-24-0154-PPD",
    doi = "10.1088/1475-7516/2025/02/021",
    journal = "JCAP",
    volume = "02",
    pages = "021",
    year = "2025"
}

@article{DESI2024BAO,
  author = "M. Abdul-Karim et al.",
    collaboration = "DESI",
    title = "{DESI DR2 results. II. Measurements of baryon acoustic oscillations and cosmological constraints}",
    eprint = "2503.14738",
    archivePrefix = "arXiv",
    primaryClass = "astro-ph.CO",
    doi = "10.1103/tr6y-kpc6",
    journal = "Phys. Rev. D",
    volume = "112",
    pages = "083515",
    year = "2025"
}

@article{Linder2005,
  author = {Linder, E. V.},
  title = {Cosmic growth history and expansion history},
  journal = {Phys. Rev. D},
  volume = {72},
  pages = {043529},
  year = {2005}
}

@article{Wang2008,
  author = {Wang, Y.},
  title = {Figure of Merit for Dark Energy Constraints from Current Observational Data},
  journal = {Phys. Rev. D},
  volume = {77},
  pages = {123525},
  year = {2008}
}

@article{Nojiri:2005pu,
    author = "Nojiri, Shin'ichi and Odintsov, Sergei D.",
    title = "{Unifying phantom inflation with late-time acceleration: Scalar phantom-non-phantom transition model and generalized holographic dark energy}",
    eprint = "hep-th/0506212",
    archivePrefix = "arXiv",
    doi = "10.1007/s10714-006-0301-6",
    journal = "Gen. Rel. Grav.",
    volume = "38",
    pages = "1285--1304",
    year = "2006"
}

@article{Nojiri:2017opc,
    author = "Nojiri, Shin'ichi and Odintsov, S. D.",
    title = "{Covariant Generalized Holographic Dark Energy and Accelerating Universe}",
    eprint = "1703.06372",
    archivePrefix = "arXiv",
    primaryClass = "hep-th",
    doi = "10.1140/epjc/s10052-017-5097-x",
    journal = "Eur. Phys. J. C",
    volume = "77",
    number = "8",
    pages = "528",
    year = "2017"
}

@article{Nojiri:2021iko,
    author = "Nojiri, Shin'ichi and Odintsov, Sergei D. and Paul, Tanmoy",
    title = "{Different Faces of Generalized Holographic Dark Energy}",
    eprint = "2105.08438",
    archivePrefix = "arXiv",
    primaryClass = "gr-qc",
    doi = "10.3390/sym13060928",
    journal = "Symmetry",
    volume = "13",
    number = "6",
    pages = "928",
    year = "2021"
}

@article{Nojiri:2021jxf,
    author = "Nojiri, Shin'ichi and Odintsov, Sergei D. and Paul, Tanmoy",
    title = "{Barrow entropic dark energy: A member of generalized holographic dark energy family}",
    eprint = "2112.10159",
    archivePrefix = "arXiv",
    primaryClass = "gr-qc",
    doi = "10.1016/j.physletb.2021.136844",
    journal = "Phys. Lett. B",
    volume = "825",
    pages = "136844",
    year = "2022"
}

@article{Nojiri:2022aof,
    author = "Nojiri, Shin'ichi and Odintsov, Sergei D. and Faraoni, Valerio",
    title = "{From nonextensive statistics and black hole entropy to the holographic dark universe}",
    eprint = "2201.02424",
    archivePrefix = "arXiv",
    primaryClass = "gr-qc",
    doi = "10.1103/PhysRevD.105.044042",
    journal = "Phys. Rev. D",
    volume = "105",
    number = "4",
    pages = "044042",
    year = "2022"
}

@article{Nojiri:2022dkr,
    author = "Nojiri, Shin'ichi and Odintsov, Sergei D. and Paul, Tanmoy",
    title = "{Early and late universe holographic cosmology from a new generalized entropy}",
    eprint = "2205.08876",
    archivePrefix = "arXiv",
    primaryClass = "gr-qc",
    doi = "10.1016/j.physletb.2022.137189",
    journal = "Phys. Lett. B",
    volume = "831",
    pages = "137189",
    year = "2022"
}

@article{Nojiri:2023bom,
    author = "Nojiri, Shin'ichi and Odintsov, Sergei D. and Paul, Tanmoy",
    title = "{Microscopic interpretation of generalized entropy}",
    eprint = "2311.03848",
    archivePrefix = "arXiv",
    primaryClass = "gr-qc",
    doi = "10.1016/j.physletb.2023.138321",
    journal = "Phys. Lett. B",
    volume = "847",
    pages = "138321",
    year = "2023"
}

@article{Hernandez-Almada:2021aiw,
    author = "Hern{\'a}ndez-Almada, A. and Leon, Genly and Maga{\~n}a, Juan and Garc{\'\i}a-Aspeitia, Miguel A. and Motta, V. and Saridakis, Emmanuel N. and Yesmakhanova, Kuralay",
    title = "{Kaniadakis-holographic dark energy: observational constraints and global dynamics}",
    eprint = "2111.00558",
    archivePrefix = "arXiv",
    primaryClass = "astro-ph.CO",
    doi = "10.1093/mnras/stac255",
    journal = "Mon. Not. Roy. Astron. Soc.",
    volume = "511",
    number = "3",
    pages = "4147--4158",
    year = "2022"
}

@article{Luciano:2026eiy,
    author = "Luciano, G. G. and Saridakis, E. N.",
    title = "{Holographic dark energy from a new two-parameter entropic functional}",
    eprint = "2603.10964",
    archivePrefix = "arXiv",
    primaryClass = "gr-qc",
    doi = "10.1016/j.physletb.2026.140674",
    journal = "Phys. Lett. B",
    volume = "879",
    pages = "140674",
    year = "2026"
}

@article{Leizerovich:2026pfy,
    author = "Leizerovich, Mat{\'\i}as and Landau, Susana J. and Luciano, Giuseppe Gaetano and Papatriantafyllou, Andreas and Saridakis, Emmanuel N.",
    title = "{Observational constraints on Luciano-Saridakis entropic cosmology}",
    eprint = "2603.03568",
    archivePrefix = "arXiv",
    primaryClass = "astro-ph.CO",
    doi = "10.1016/j.dark.2026.102333",
    journal = "Phys. Dark Univ.",
    volume = "52",
    pages = "102333",
    year = "2026"
}

@article{Luciano:2025ykr,
    author = "Luciano, Giuseppe Gaetano and Paliathanasis, Andronikos",
    title = "{Late-time cosmological constraints on Kaniadakis holographic dark energy}",
    eprint = "2509.17527",
    archivePrefix = "arXiv",
    primaryClass = "astro-ph.CO",
    doi = "10.1140/epjc/s10052-025-15122-9",
    journal = "Eur. Phys. J. C",
    volume = "85",
    number = "12",
    pages = "1384",
    year = "2025"
}

@article{Luciano:2025hjn,
    author = "Luciano, Giuseppe Gaetano and Paliathanasis, Andronikos and Saridakis, Emmanuel N.",
    title = "{Barrow and Tsallis entropies after the DESI DR2 BAO data}",
    eprint = "2504.12205",
    archivePrefix = "arXiv",
    primaryClass = "gr-qc",
    doi = "10.1088/1475-7516/2025/09/013",
    journal = "JCAP",
    volume = "09",
    pages = "013",
    year = "2025"
}

@article{Cohen:1998zx,
    author = "Cohen, Andrew G. and Kaplan, David B. and Nelson, Ann E.",
    title = "{Effective field theory, black holes, and the cosmological constant}",
    eprint = "hep-th/9803132",
    archivePrefix = "arXiv",
    reportNumber = "BUHEP-98-7, DOE-ER-40561-358, INT-98-00-6, UW-PT-97-24",
    doi = "10.1103/PhysRevLett.82.4971",
    journal = "Phys. Rev. Lett.",
    volume = "82",
    pages = "4971--4974",
    year = "1999"
}

@article{Trotta:2008qt,
    author = "Trotta, Roberto",
    title = "{Bayes in the sky: Bayesian inference and model selection in cosmology}",
    eprint = "0803.4089",
    archivePrefix = "arXiv",
    primaryClass = "astro-ph",
    doi = "10.1080/00107510802066753",
    journal = "Contemp. Phys.",
    volume = "49",
    pages = "71--104",
    year = "2008"
}

@article{Liddle:2007fy,
    author = "Liddle, Andrew R",
    title = "{Information criteria for astrophysical model selection}",
    eprint = "astro-ph/0701113",
    archivePrefix = "arXiv",
    doi = "10.1111/j.1745-3933.2007.00306.x",
    journal = "Mon. Not. Roy. Astron. Soc.",
    volume = "377",
    pages = "L74--L78",
    year = "2007"
}

@article{Ong:2025utx,
    author = "Ong, Dily Duan Yi and Yallup, David and Handley, Will",
    title = "{A Bayesian Perspective on Evidence for Evolving Dark Energy}",
    eprint = "2511.10631",
    archivePrefix = "arXiv",
    primaryClass = "astro-ph.CO",
    month = "11",
    year = "2025"
}
\bibliographystyle{elsarticle-num}

\end{document}